\documentclass[
    aps,
    prx,
    superscriptaddress,
    twocolumn,
    a4paper,
    floatfix,
    longbibliography
]{revtex4-2}

\usepackage[american]{babel}
\usepackage[utf8x]{inputenc}

\usepackage{grffile} 

\usepackage{newtxtext,newtxmath}
\usepackage{microtype}

\usepackage{amsmath}
\usepackage{amssymb}
\usepackage{bbm}
\usepackage{mathtools}
\usepackage{dsfont}
\usepackage{braket}
\usepackage{cancel}
\usepackage{slashed}

\usepackage{graphicx}
\usepackage[svgnames]{xcolor}

\usepackage{siunitx} 

\usepackage{ragged2e}
\usepackage{array}
\usepackage{tabularx}
\usepackage{booktabs}

\definecolor{cset-aps-blueberry}{RGB}{28,128,158}
\definecolor{cset-aps-blue}{RGB}{46,44,184}
\definecolor{cset-aps-turquoise}{RGB}{0,67,88}
\definecolor{cset-aps-limegreen}{RGB}{190,219,67}
\definecolor{cset-aps-green}{RGB}{31,138,112}
\definecolor{cset-aps-yellow}{RGB}{255,225,25}
\definecolor{cset-aps-orange}{RGB}{253,116,0}
\definecolor{cset-aps-red}{RGB}{219,0,43}

\usepackage{tikz}

\usepackage{pgfplots}
\pgfplotsset{%
    every axis legend/.append style={%
        cells={anchor=west},
        at={(0.96,0.04)},
        anchor=south east,
        font=\scriptsize,
        },
    every axis/.append style={%
        yticklabel style={%
            /pgf/number format/fixed zerofill,
            /pgf/number format/precision=2},
        },
    width= \textwidth,
    height=8cm,
    xmajorgrids=true,
    xminorgrids=false,
    minor x tick num=1,
}
\usepgfplotslibrary{external}

\usetikzlibrary{decorations}

\usepackage{pict2e,picture}

\makeatletter
\DeclareRobustCommand{\Arrow}[1][]{%
\check@mathfonts
\if\relax\detokenize{#1}\relax
\settowidth{\dimen@}{$\m@th\rightarrow$}%
\else
\setlength{\dimen@}{#1}%
\fi
\sbox\z@{\usefont{U}{lasy}{m}{n}\symbol{41}}%
\begin{picture}(\dimen@,\ht\z@)
\roundcap
\put(\dimexpr\dimen@-.7\wd\z@,0){\usebox\z@}
\put(0,\fontdimen22\textfont2){\line(1,0){\dimen@}}
\end{picture}%
}
\makeatother
\newcommand{\veryshortrightarrow}{\hspace{.2mm}\scalebox{.8}{\Arrow[.1cm]}\hspace{.2mm}}

\usepackage{hyperref}
\hypersetup{%
    colorlinks=true,
    linkcolor={cset-aps-red},
    linkbordercolor={cset-aps-red},
    filecolor={cset-aps-orange},
    filebordercolor={cset-aps-orange},
    citecolor={cset-aps-blue},
    citebordercolor={cset-aps-blue},
    urlcolor={cset-aps-green},
    urlbordercolor={cset-aps-green},
    menucolor={cset-aps-limegreen},
    menubordercolor={cset-aps-limegreen},
    breaklinks=true,
    pdfborderstyle={/S/U/W 2},
    pdfpagemode=UseOutlines,
    pdfstartpage={1},
}

\newcommand{\ii}{\text{i}}

\usepackage{lipsum}

\begin{document}

\title[Gravitational Redshift Tests with Atomic Clocks and Atom Interferometers]{Gravitational Redshift Tests with Atomic Clocks and Atom Interferometers}
\collaboration{This article has been published in \href{https://journals.aps.org/prxquantum/abstract/10.1103/PRXQuantum.2.040333}{PRX Quantum \textbf{2}, 040333 [2021]}}
\author{Fabio Di Pumpo}
\email{fabio.di-pumpo@uni-ulm.de}
\author{Christian Ufrecht}
\author{Alexander Friedrich}
\address{Institut f{\"u}r Quantenphysik and Center for Integrated Quantum
    Science and Technology (IQ\textsuperscript{ST}), Universit{\"a}t Ulm, Albert-Einstein-Allee 11, D-89069 Ulm, Germany}
\author{Enno Giese}
\address{Institut f{\"u}r Quantenphysik and Center for Integrated Quantum Science and Technology (IQ\textsuperscript{ST}), Universit{\"a}t Ulm, Albert-Einstein-Allee 11, D-89069 Ulm, Germany}
\address{Institut f{\"u}r Angewandte Physik, Technische Universit{\"a}t Darmstadt, Schlossgartenstr. 7, D-64289 Darmstadt, Germany}
\address{Institut f{\"u}r Quantenoptik, Leibniz Universit{\"a}t Hannover, Welfengarten 1, D-30167 Hannover, Germany}
\author{Wolfgang P. Schleich}
\address{Institut f{\"u}r Quantenphysik and Center for Integrated Quantum Science and Technology (IQ\textsuperscript{ST}), Universit{\"a}t Ulm, Albert-Einstein-Allee 11, D-89069 Ulm, Germany}
\address{Institute of Quantum Technologies, German Aerospace Center (DLR), S\"{o}flinger Stra\ss e 100, D-89077 Ulm, Germany}
\address{Hagler Institute for Advanced Study and Department of Physics and Astronomy, Institute for Quantum Science and Engineering (IQSE), Texas A{\&}M University, College Station, Texas 77843-4242, USA}
\address{Texas A{\&}M AgriLife Research, Texas A{\&}M University, College Station, Texas 77843-4242, USA}
\author{William G. Unruh}
\address{Hagler Institute for Advanced Study and Department of Physics and Astronomy, Institute for Quantum Science and Engineering (IQSE), Texas A{\&}M University, College Station, Texas 77843-4242, USA}
\address{Department of Physics and Astronomy, University of British Columbia, Vancouver, British Columbia V6T 1Z1, Canada}

\begin{abstract}
\noindent Atomic interference experiments can probe the gravitational redshift via the internal energy splitting of atoms and thus give direct access to test the universality of the coupling between matter-energy and gravity at different spacetime points.
By including possible violations of the equivalence principle in a fully quantized treatment of all atomic degrees of freedom, we characterize how the sensitivity to gravitational redshift violations arises in atomic clocks and atom interferometers, as well as their underlying limitations.
Specifically, we show that:
(i.)~Contributions beyond linear order to trapping potentials lead to such a sensitivity of trapped atomic clocks.
(ii.)~While Bragg-type interferometers, even with a superposition of internal states, with state-independent, linear interaction potentials are at first insensitive to gravitational redshift tests, modified configurations, for example by relaunching the atoms, can mimic such tests tests under certain conditions and may constitute a competitive alternative.
(iii.)~Guided atom interferometers are comparable to atomic clocks.
(iv.)~Internal transitions lead to state-dependent interaction potentials through which light-pulse atom interferometers can become sensitive to gravitational redshift violations.
\end{abstract}

\maketitle

\section{Introduction}
\label{sec.Intr}
The phase of a matter wave is connected to proper time~\cite{Storey1994}. 
Therefore, at first sight atom interferometers, which compare phases accumulated along different branches, seem to be a natural candidate to probe the universality of the gravitational redshift (UGR)~\cite{Mueller2010,Hohensee2011,Lan2013}. 
This foundational principle of general relativity can be tested by comparing relative phases of two independent (atomic) clocks at different heights, measuring the proper-time differences between them.
However, the analogy to conventional light-pulse atom interferometers is misleading~\cite{Wolf2010,Mueller2010b,Unnikrishnan2011,Schleich2013}, as they usually lack the internal energy difference that leads to a physical clock~\cite{Wolf2011,Hohensee2012,Wolf2012,Sinha2011,Zych2011,Peil2014}.
Despite this objection, recent proposals based on generating superpositions during the interferometer~\cite{Roura2020} or relying on internal transitions~\cite{Ufrecht2020} are indeed able to test UGR.
In this article, we identify in analogy to atomic clocks the reason for a sensitivity of atom interferometers to UGR violations.
To this end, we derive a generalized treatment for both atomic clocks and atom interferometers including physics beyond the Standard Model by considering non-Einsteinian components of gravity, establishing a common and rigid criterion for being sensitive to UGR violations.

In its operational definition, proper time is the quantity measured by an idealized clock on a given worldline and manifests itself---as a direct consequence of special and general relativity---in time dilation between two observers~\cite{Einstein1905,Einstein1907,Einstein1911}.
As such, time dilation can be used to test a basic premise of general relativity, the Einstein equivalence principle (EEP), which translates into three core assumptions~\cite{DiCasola2015}: local Lorentz invariance, universality of free fall (UFF), and universality of the gravitational redshift.
In principle, UGR is a special case of local position invariance and tests the expected gravitational redshift measured between two identical clocks propagating along different worldlines~\cite{Giulini2012}.
Alternative tests of local position invariance are verifications of the universality of the rates of two different clocks on the same worldline~\cite{Will2014}.
Since we focus in our article on the comparison with atom interferometers and thus on different worldlines, we do not consider this kind of test.
If and only if EEP is valid, gravity can be described as a metric theory.
In such theories the metric solely contains all information on the motion of massive test bodies in gravity without any additional field or physical object~\cite{Will1974,Horvath1988,Damour1996}.
Thus, EEP demands that gravity couples in a universal manner, i.\,e., composition independently, to all types of non-gravitational matter-energy~\cite{Will2014}. 
However, even further assumptions are necessary to refine the variety of metric theories to the theory of general relativity~\cite{DiCasola2015}.

Atom interferometers routinely examine in ambitious UFF tests~\cite{Schlippert2014,Asenbam2020} whether the free fall of a test particle depends on the composition of its rest mass.
Hence, it seems natural and worthwhile to study their suitability for tests of UGR.
A widely accepted operational definition of violations of UGR is derived from atomic clocks. 
Employing this definition, we show in the present article that every conventional Bragg-type atom interferometer~\cite{Giese2015} in a uniform gravitational field with only a single internal state of the atom does not test UGR if it is closed, which means that there is no separation in position and momentum between both branches at the readout time.
A Bragg-type interferometer does not change the internal state, so that the employed diffraction is described by linear interaction potentials that do not depend on the internal state of the atom.
However, for closed Bragg-type atom interferometers operated with a superposition of internal states of the atom, we analyze under which conditions such schemes can mimic UGR tests akin to atomic clocks.
These results are surprising, given that conventional light-pulse atom interferometers do not test UGR~\cite{Loriani2019}.
Hence, we discuss the impact of such schemes on future experiments using a specific geometry and show that they can push the current limits of UGR tests with atom interferometry in free fall.
Moreover, we demonstrate how other types of atom interferometers can be sensitive to UGR violations. 
The latter include guided atom interferometers or interferometers employing linear but state-dependent interactions, for example through internal transitions during the sequence.

\subsection{Distinction between UGR and UFF}

Under the condition of energy conservation, UGR can be seen as a consequence of attributing weight to matter-energy~\cite{Unruh1979}.
It assumes that the ticking rate of an idealized clock is altered in the presence of a gravitational field, and that the magnitude of this (universal) modification must not depend on the composition of the clock. 
The phase accumulated by a matter wave is proportional to the mass of the particle including also contributions from the internal energy.
Therefore, it was argued that the associated Compton frequency represents the ticking rate of a clock~\cite{Mueller2010}, even if the experiment is performed with only one internal state of the atom.
This claim was opposed (see, for example, Ref.~\cite{Wolf2010}), because UGR tests are conventionally parametrized by violation models coupling to different internal states of an atom. 
As such, UGR tests compare the local ticking rate determined by the internal degrees of freedom of an atomic clock.
Consequently, a consistent parametrization of UGR violations has to involve at least two internal states.

In contrast, UFF tests are usually performed through a comparison (null test) of the gravitational accelerations of
\emph{different and independent test masses}.
However, it is also possible to compare the acceleration experienced by an atom in different \emph{internal states}~\cite{Zhang2018}.
Because the test object is the same for both internal states, UFF violations between different internal states are suppressed compared to UFF violations with different species~\cite{Damour1999}.
In these situations, however, the violation parameters of UFF and UGR can be connected on the basis of fundamental assumptions such as energy conservation~\cite{Lightman1973,Nordtvedt1975,Haugan1979}.
Assuming that only such EEP violations from different internal states are considered, there are two fundamental differences between UFF and UGR tests, which allow for a consistent distinction between both principles:
(i)~Even though the UFF and UGR violation parameters are not independent, they couple differently in experimental situations.
While UFF violations are connected to the center-of-mass dynamics of the total mass-energy of the test object, UGR violations depend on the internal energy difference.
Hence, depending on the specific situation, the ratio of these two energy scales can either enhance or suppress the respective violation parameter.
(ii)~UFF tests are by definition null tests of two gravitational accelerations.
In contrast, tests of UGR probe the universality (composition independence) of a nonvanishing gravitational effect~\cite{Will2014} between different worldlines as a consequence of local position invariance.
A comparison of physical systems with different compositions on the same worldline constitutes a null test of local position invariance~\cite{Peil2013,Ashby2018}.
However, since we focus in our article on the application to atom interferometers that intrinsically include different worldlines, we do not consider this kind of test.

\textit{A priori} UGR and UFF constitute two distinct premises of metric theories of gravity~\cite{Will2014,DiCasola2015}; and the interdependence of UFF and UGR in our framework does not necessarily imply that they are as well connected in a still unknown fundamental theory.
On a methodological level, conducting tests of both principles independently is key to a complementary approach with minimal assumptions and conjectures~\cite{Wolf2016}.
However, already within violation models where both principles are related~\cite{Lightman1973}, such as ours, it is important to distinguish between the sensitivity to different types of tests.
For example, the sensitivity to EEP violations in UGR tests or in UFF tests depends crucially on the coupling strengths of the violation models to the involved forces and particles, and the exact compositions of the test particles.
Consequently, there are scenarios where UGR tests are much more sensitive than UFF tests and vice versa~\cite{Wolf2011}.

Our article focuses on UGR.
This principle can be tested through transitions of nuclear resonance experiments~\cite{Pound1959,Pound1960} as well as through atomic clocks moving on different worldlines~\cite{Opat1991} subject to kinetic and gravitational time dilation.
Such tests have been performed aboard airplanes~\cite{Hafele1972a,Hafele1972b} and space crafts~\cite{Kleppner1970,Vessot1980,Delva2015,Herrmann2018,Delva2018,Delva2019,Savalle2019}.
Today's state-of-the-art (optical) clocks~\cite{Nicholson2015,Brewer2019,Oelker2019,Madjarov2019} offer unprecedented accuracy of time keeping and accurate UGR tests~\cite{Takamoto2020} on small distances~\cite{Chou2010,Bothwell2021,Zheng2021}.

\subsection{Phases of matter waves}
All of these classical clock schemes rely on the comparison of two independent physical systems, for example two locally trapped atoms forming atomic clocks.
However, because the phase of a matter wave is intrinsically connected to proper time, one could naively think that a single atom in spatial superposition can also be used for UGR tests.
Indeed, in the low-velocity and weak-field approximation the global phase of a first-quantized matter wave 
\begin{equation}
\label{eq:PhaseOfMatterWave}
    \phi=S/\hbar=-\int\!\mathrm{d}t\,\frac{m}{\hbar}\left[c^2-\frac{\dot{z}^2}{2}+ U\left(z\right)+ \frac{V\left(z,t\right)}{m}\right]
\end{equation}
without any violation of EPP depends on the action $S$ of a classical point particle with mass $m$ and velocity $\dot{z}$ including the Newtonian gravitational field $U(z)$.
It is obtained from the expansion of a point particle's proper time up to $c^{-2}$, parametrized through the lab time $t$.
The additional potential $V\left(z,t\right)$ contains the interactions that prevent the matter from following a geodesic.
The interactions can, for example, originate from a trap, a guiding potential, or light pulses.
The mass includes internal-energy contributions so that possible internal transitions during the experiment imply a time-dependent mass.
Because of the unidimensional configurations considered in this article, only one spatial direction suffices for our description.

Since the classical action $S$ is encoded as a global phase, it is impossible to obtain the value of this phase from a quantum mechanical experiment.
Instead, one must interfere two systems (or one system in two configurations).
In fact, an atom interferometer brings an atom into a superposition of two different worldlines and allows for a measurement of the action difference between the interferometer branches through the relative \emph{interferometric} phase $\varphi$.
Since this phase does in principle include relativistic effects~\cite{Dimopoulos2008,Tan2017}, it seems plausible that such an experiment is sufficient for tests of UGR.
However, a conventional Bragg-type atom interferometer is operated only in one internal state without internal transitions.
Therefore, it lacks an energy reference necessary for a periodic system that constitutes a clock, so that it is impossible to detect UGR violations.
To introduce such a quantum clock~\cite{Sinha2011,Zych2011,[][ and references therein.]Pikovski2017,Loriani2019}, an atom interferometer can bring an initial superposition of two internal states (similar to a Ramsey sequence of atomic clocks) into a spatial superposition.
This concept of quantum clock interferometry raises the question whether tests of UGR are possible in such configurations and of their underlying principles.

\subsection{Outline and key results}
As an example of a theory that contains violations of UFF and UGR, we rely in this article on a dilaton model~\cite{Alves2000,Damour1999,Damour2010,Damour2012,Roura2020}, discussed in Sec.~\ref{sec.violations_dilaton} and further detailed in Appendix~\ref{sec:DilMod}.
We introduce in Sec.~\ref{sec.perturbations} a common formalism for interferometric atomic clocks as well as for atom interferometers and identify the crucial mechanisms for UGR tests in both interferometric experiments.
In Sec.~\ref{sec.UGR_with_atomic_clocks} we observe that the sensitivity of atomic clocks to tests of UGR is based on the nonlinear position dependence of the trapping potentials that force both internal states to follow a common trajectory. 
Consequently, the differential phase between two clocks is distinguishable from a situation in a freely falling frame, even if there is no violation of the EEP.

In contrast, we demonstrate in Sec.~\ref{sec.UGR_with_QCI} that all closed Bragg-type atom interferometers with only a single internal state of the atom are fundamentally insensitive to violations of the gravitational redshift in a linear gravitational field.
Even if operated as a quantum clock, i.e., with an internal superposition, such interferometer experiments are at first different from UGR tests performed with atomic clocks.
As such, a Mach-Zehnder configuration is the prime example for UFF tests, even though there is a fundamental connection between our parametrization of UFF violations and those associated with UGR.
However, we show that modifications of Bragg-type interferometers are possible, for example by applying relaunch pulses, so that they mimic UGR tests akin to atomic clocks under certain conditions.
Besides, we identify a large class of other UGR-sensitive atom interferometer schemes:
if the internal state is not changed by the light-matter interaction, UGR tests can be achieved---in analogy to clocks---through potentials that are nonlinear in the atom's position, forcing both internal states on one common trajectory (guiding schemes).
Alternatively, Sec.~\ref{sec.UGR_variable_mass} shows that changing the internal state of the atom during the light-pulse sequence of an atom interferometer~\cite{Roura2020,Ufrecht2020} generates a state-dependent momentum transfer that may result in a UGR sensitivity.

We put our results in Sec.~\ref{sec.Discussion} into perspective, discuss possible implementations, and give rough estimates to assess the influence of the discussed schemes on future experimental realizations.

\section{Violations caused by dilaton fields}
\label{sec.violations_dilaton}
In the following we introduce mass defects caused by different internal energies as well as an EEP-violation model based on the coupling to a dilaton field. 
We are then able to discuss EEP violations detected by idealized clocks and use exactly such a setup to identify and define violations of UGR.

\subsection{Mass defects and dilaton coupling}
The idea of internal superpositions as an input for atom interferometers~\cite{Sinha2011,Zych2011,Rosi2017} triggered interest in the consistent description of mass defects caused by the different internal energies~\cite{Jaervinen2005,Gooding2015,Pikovski2017,Anastopoulos18,Sonnleitner2018,Schwartz2019}.
Similarly, we introduce relativistic corrections through the internal energy $E_j\equiv m_jc^2$ of two internal states $j=a,b$.
Generalizing Eq.~\eqref{eq:PhaseOfMatterWave}, this relation implies a state-dependent mass $m_{b/a}=m\pm\Delta m/2$ and leads to a different action $S_j$ for each internal state.
At this point it is interesting to note that solely the assumption that energy possesses weight directly implies the gravitational redshift~\cite{Unruh1979}.

To introduce deviations from the established laws of gravitation, one possibility is to consider a massless, scalar dilaton field. 
The coupling of this field to gravity, the elementary particles, and the gauge fields, in particular the electromagnetic one, arises from an effective action as a result, for example, of string theory~\cite{Damour1994}.
After a conformal transformation, modified Einstein field equations emerge, together with a dilaton coupling to all elementary particles and all other forces of the Standard Model.
By linearizing the interacting theory in the dilaton field, we obtain an effective position-dependent interaction between dilaton and all other fields.
In the low-energy expansion in orders of $c^{-1}$ this interaction alters the atom's mass (including internal energies)~\cite{Damour1999,Damour2010,Damour2012,Roura2020} and by that introduces a nonuniversal coupling to gravity~(see Appendix~\ref{sec:DilMod}).
After approximating the theory with a first-quantized description, we expand the atom's dilaton-dependent mass (including the mass defect through different internal energies) to linear order in the dilaton field around its Standard-Model value. 
This way, we obtain $m_j\big[1+\beta_jU\left(z\right)/c^2\big]$ including the Newtonian gravitational field $U(z)$.
The dimensionless parameter $\beta_j$ that serves as the EEP-violation parameter is the linear expansion coefficient.
In this order in $c^{-2}$, the interaction of the atom with a light field is not influenced by the dilaton field.
In Appendix~\ref{sec:DilMod} we present a more detailed model from a field-theoretical perspective to motivate this relation.

However, both UGR and UFF violations are connected in the dilaton model and can be exactly parametrized in this specific violation framework.
This connection relies on energy conservation. 

\subsection{UGR violations in idealized clocks}
An idealized clock in general relativity is operationally defined as a physical system that measures proper time along a worldline.
This clock hypothesis~\cite{Einstein1911} determines the mathematical expression for proper time.
Since in the presence of a dilaton field the coupling of gravity to a physical system that constitutes a clock depends on the nonuniversal parameter $\beta_j$, issues arise in such an operational definition.
However, one can still introduce an idealized clock, consisting of a superposition of two internal states of an atom. 
This clock measures general-relativistic proper time together with an additional nonuniversal contribution caused by the dilaton field.

In accordance with this definition, we introduce an internal structure to the system and assume that such an idealized clock consists of a superposition of two colocalized internal states (masses) moving along one common trajectory.
However, atomic clocks are based on internal transitions generated by the absorption or emission of photons.
As a result the associated photon recoil has profound consequences on the limitations of measurements of spacetime distances~\cite{Salecker1958}.
To circumvent such issues, we assume recoilless internal transitions or work in a regime where recoil effects are negligible.
We study the effect of transferred momentum through the interaction with light gratings in the context of atom interferometers, where this effect is the main mechanism to generate spatial superpositions.

For atomic clocks, the dilaton field causes the two internal states to drop at different rates and by that to follow different trajectories during the operation of the clock.
To mitigate this effect, we assume an idealized state-independent trapping potential $V$ that is sufficiently steep to guide both masses on a common trajectory $\zeta(t)$.
We extend the model in Sec.~\ref{sec.UGR_with_atomic_clocks} to clocks with quantized center-of-mass (c.m.) motion and nonidealized trapping potentials.
The action from Eq.~\eqref{eq:PhaseOfMatterWave} for time-independent masses, including the mass defect $m_{b/a}=m\pm\Delta m/2$, is modified by the dilaton field leading to $S_j(\zeta)-m_j\beta_j\int\!\mathrm{d}t\,U(\zeta)$.
Here, we used the fact that the EEP violation $\beta_j$ effectively leads to a state-dependent gravitational potential $(1+\beta_j)U$.
With this modified action, we find the phase
\begin{equation}
    \label{eq:PhaseIdealSingleClock1}
    \varphi=\frac{1}{\hbar}\left\lbrace \left.\left(S_b-S_a\right)\right|_{\zeta}-m\,\Delta\beta\!\!\int\!\!\mathrm{d}t\,U(\zeta)\right\rbrace
\end{equation}
between both masses measured by a single idealized clock on one trajectory $\zeta(t)$.
Here, we introduced the parameter $\Delta\beta=\left(\beta_b-\beta_a\right)$ and neglected $\Delta m\,\beta_j$~\footnote{In fact, keeping these terms leads to a redefinition of the gravitational potential, which effectively corresponds to using $m$ as a test mass for its determination.}.
Moreover, we assume the contribution to the action from the idealized potential $V$ to be state independent.
It thus cancels in the phase and we can focus on the contribution $S_j(\zeta)=-m_j\!\int\!\mathrm{d}t\,\left[c^2-\dot{\zeta}^2/2+ U(\zeta)\right]$ from the Newtonian part of the action~\footnote{In principle, laser phases are imprinted during the transition, but can be trivially added to the description.
Moreover, they may cancel in differential schemes that are used to measure UGR violations}.
In the Newtonian case ($\Delta\beta=0$), this idealized clock therefore measures the relative phase $(S_b-S_a)/\hbar=-\Omega\tau$ between the two internal states with frequency $\Omega=\Delta m c^2/\hbar$ and proper time $\tau\cong \int\!\mathrm{d}t~\left[1-\dot{\zeta}^2/\big(2c^2\big)+U(\zeta)/c^2\right]$ along the trajectory $\zeta(t)$ of the trap.
Including the dilaton field, the phase from Eq.~\eqref{eq:PhaseIdealSingleClock1} takes the form
\begin{equation}
    \label{eq:PhaseIdealSingleClock2}
    \varphi=-\Omega\left[\tau+\alpha\!\int\!\mathrm{d}t\,U(\zeta)/c^2\right],
\end{equation}
where we have introduced the parameter $\alpha=m\Delta\beta/\Delta m$, which parametrizes UGR violations.
Conversely, we show that UFF violations of internal states are parametrized through $\Delta\beta$ in this framework.
Thus, $\alpha$ highlights the fundamental connection between both principles.
This relation can be proven by general energy-conservation arguments~\cite{Nordtvedt1975,Haugan1979} even for a wider range of nonmetric theories beyond the dilaton model.

To identify UGR violations, it is necessary to determine the differential phase between two clocks moving along different trajectories (in our violation model) and then compare it to the classical expression for the proper time of ideal clocks moving along trajectories with the same initial conditions ticking at rate $\Omega$.
When we introduce two such branches, the phase $\varphi^{(\sigma)}$ becomes branch dependent, so that it has to be denoted by a superscript $\sigma=u,l$ for an upper and lower branch.
To simplify the discussion, we consider from here on only linear Newtonian potentials $U\left(z\right)=g z$ with gravitational acceleration $g$.
We then find with Eq.~\eqref{eq:PhaseIdealSingleClock2} for two stationary traps separated by a fixed distance $\delta\zeta_0=\zeta^{(u)}-\zeta^{(l)}$ the differential phase
\begin{equation}
    \label{eq:PhaseIdealClockOnTwoHeights}
    \Phi_\text{C}=\varphi^{(u)}-\varphi^{(l)}=-\Omega\left(1+\alpha\right)\delta\zeta_0 g T/c^2
\end{equation}
between the two clocks, where $u$ and $l$ denote the upper and lower branches.
The proper-time difference between the two trajectories is caused by gravitational time dilation and given by $\delta\tau=\delta\zeta_0 g T/c^2$, where $T$ is the time between simultaneous initialization and readout in the laboratory frame.
Here, we neglected finite speed-of-light effects for the moment.
Finally, Eq.~\eqref{eq:PhaseIdealClockOnTwoHeights} defines a classical UGR test measuring violations $\alpha$ that modify unity as a prefactor and depends on the internal composition of the atoms.
Note that, even though $\alpha = m \Delta \beta / \Delta m $ is connected to violations of UFF parametrized by $\Delta\beta$, idealized clocks test a modification of the gravitational redshift by the factor $(1+\alpha)$.

\section{Clocks with quantized motion and quantum clock interference}
In this section we first present a perturbative formalism for the calculation of interferometric phases suitable for describing both atomic clocks and atom interferometers.
We then apply the method to two atomic clocks and show that different schemes comparing their phases lead to classical UGR tests.
In the final part of this section, we transfer the concept to atom interferometers.
While they are in the first place insensitive to UGR violations in analogy to atomic clocks, we show that they can be used under certain conditions to mimic UGR tests akin to those performed with atomic clocks. 

\subsection{Interferometric phases from perturbations}
\label{sec.perturbations}
Atomic clocks in a Ramsey sequence are based on the interference of two internal states, whereas, in atom interferometers, the two branches interfere c.m. wave packets.
Hence, both types of experiments rely on the interference of two components $\ket{\psi_1}$ and $\ket{\psi_2}$.
The form of these two components depends on the observable, i.e., a projector, that defines the exit port of the interference experiment.
After the projection associated with the observable, we find that the component $\ket{\psi_{1,2}}= \hat{U}_{1,2} \ket{\psi_\text{in}}/2$ arises from the action of an effective evolution operator $\hat{U}_{1,2}$ on the initial state $\ket{\psi_\text{in}}$; see Appendix~\ref{sec:interference_signal}. 
For atomic clocks, the population of one of the two internal states is the observable of interest, for example measured by fluorescence.
In contrast, the exit port of atom interferometers is defined through specific values of the atom's final momentum that can be measured by absorption imaging in the far field.
Both operators $\hat{U}^{}_1$ and $\hat{U}^{}_2$ incorporate the projection onto the exit port and can be associated with different effective time evolutions and corresponding effective Hamiltonians that describe the generation of the individual components dynamically.
In this case, the interference pattern is determined by the expectation value of the overlap $\big<\hat{U}^{\dagger}_1\hat{U}^{}_2\big>=C \exp({\ii\varphi})$ with respect to the initial state. 
Here, $C$ denotes the contrast of the interferometer and $\varphi$ the measured interferometric phase.
Since we are interested in the effects of the c.m. motion of clocks, we incorporate the generation of internal superpositions in the initial state and the readout through a projection on the superposition of internal states.
Hence, for clocks, the operators $\hat{U}^{}_1$ and $\hat{U}^{}_2$ describe the c.m. motion of different internal states.
In contrast, for a description of atom interferometers, these operators propagate an atom in a superposition of two spatially separated branches generated by diffraction from light fields.
In both cases, the expectation value $\big<\hat{U}^{\dagger}_1\hat{U}^{}_2\big>$ is taken with respect to the initial state of the c.m. motion.

The mass defect as well as the dilaton field cause slightly different trajectories and actions compared to the original, unperturbed expressions $z(t)$ and $S$ from Eq.~\eqref{eq:PhaseOfMatterWave}.
However, the leading-order corrections to the phase are obtained by integrating the perturbation along the unperturbed trajectories~(see~Appendix~\ref{sec:ClassEqMot}).
A more rigorous justification of this intuition for operator-valued expressions and generalization to higher orders in the perturbation can be found in Refs.~\cite{Ufrecht2019,Ufrecht20202}.
Therefore, contributions from corrections to the trajectories would be additionally suppressed.
Hence, we find that the phase
\begin{equation}
    \label{eq:GeneralPhasePerturb}
    \varphi=\varphi_0-\frac{1}{\hbar}\int\!\text{d}t\,\mathcal{H}_\text{diff}+\varphi_\text{WP}
\end{equation}
calculated in a laboratory frame with laboratory time $t$ can be expressed as a sum of three contributions:
(i) the phase $\varphi_0$ measured by the closed, unperturbed interferometer that is connected to the nonrelativistic action difference~\cite{Storey1994};
(ii) a term $\mathcal{H}_\text{diff}$ originating  from the difference between two branches or two internal states of a perturbing Hamiltonian $\hat{\mathcal{H}}$ evaluated at the unperturbed phase-space trajectories;
and (iii) a contribution $\varphi_\text{WP}$ caused by wave-packet effects.
These effects arise if the two components of the wave function have obtained different shapes during the evolution and are described in more detail in Appendix~\ref{sec:WPEffects}.
The sum of the first two contributions corresponds to the action difference~\cite{Dimopoulos2008,Roura2014,Ufrecht2019}, including the lowest order of the perturbation $\hat{\mathcal{H}}_j^{(\sigma)}$. 

For the description of the c.m. motion of atomic clocks and atom interferometers without internal transitions during the sequence, we rely on an effective branch-dependent model 
\begin{equation}
    \label{eq:HamMassCorr}
    \hat{H}_j^{(\sigma)}=m c^2+\frac{\hat{p}^2}{2m}+m g\hat z+\hat{V}^{(\sigma)}+\hat{\mathcal{H}}_j^{(\sigma)}
\end{equation}
which arises after postselection and leads to the phase presented in Eq.~\eqref{eq:GeneralPhasePerturb}.
This Hamiltonian contains the expansions of proper time analogous to Eq.~\eqref{eq:PhaseOfMatterWave} up to order $c^{-2}$, the mass defect to order $\Delta m$, and the gravitational potential including the dilaton modification to linear order $(1+\beta_j)g$.
Consequently, it consists of an unperturbed part $m c^2+\hat{p}^2/(2m)+m g\hat z$ involving a linear Newtonian field together with an interaction potential
\begin{equation}
    \label{eq:NonRelPotential}
    \hat{V}^{(\sigma)}=-F^{(\sigma)}\hat z+\frac{m\Gamma^2}{2}(\hat{z}-\zeta^{(\sigma)})^2+V_\text{ph}^{(\sigma)}
\end{equation}
with branch-dependent forces $F^{(\sigma)}(t)$. 
This interaction includes time-dependent instantaneous momentum transfer and a harmonic trapping potential of frequency $\Gamma$ that is centered around the classical position $\zeta^{(\sigma)}(t)$ and is independent of $g$.
Because the harmonic trap is in general sensitive to the internal energy and thus each internal state $\ket{j}$ experiences a slightly different trapping frequency $\Gamma_j$, we have introduced the harmonic mean of the trapping potentials through $\Gamma^2=(\Gamma^2_a+\Gamma^2_b)/2$ and treat deviations as perturbations.
Diffracting laser pulses and their phases are divided into two parts:
one contribution, depending on position is included in the forces $F^{(\sigma)}$.
The position-independent part, which in principle depends on time and the branch, is contained in $V^{(\sigma)}_\text{ph}(t)$.
Because it does not contribute to the c.m. motion, we can include the phase arising from the action of $V^{(\sigma)}_\text{ph}(t)$ in the phase $\varphi_0$~\cite{Schleich2013,Loriani2019}.

Branch- and state-dependent c.m. perturbations 
are described by $\hat{\mathcal{H}}_j^{(\sigma)}$.
This contribution takes the form
\begin{align}
\begin{split}
    \label{eq:PerturPot}
    \hat{\mathcal{H}}_j^{(\sigma)}=& \lambda_j\frac{\Delta m }{2}\left[c^2-\frac{\hat{p}^2}{2 m^2}+g\hat z\right]+m\beta_j g\hat{z} \\
    &+\lambda_j\frac{m\Delta\Gamma^2}{4}(\hat{z}-\zeta^{(\sigma)})^2,
\end{split}
\end{align}
where $\lambda_b = +1$ corresponds to the exited state and $\lambda_a = -1$ to the ground state.
It contains the mass splitting $\Delta m$ that modifies the evolution in a linear gravitational field~\cite{Lammerzahl1995,Anastopoulos18,Sonnleitner2018,Schwartz2019}, the coupling through $\beta_j$ to the dilaton field~\cite{Alves2000,Damour2010,Damour2012}, and a splitting in the state-dependent frequency of the harmonic trap with $\Delta\Gamma^2=\Gamma^2_b-\Gamma^2_a$ as perturbation parameters.
Including the mass $m$ in Eqs.~\eqref{eq:NonRelPotential} and~\eqref{eq:PerturPot} as a prefactor to the harmonic potential only simplifies the notation, but does not imply that the optical trap itself is mass dependent.
In this description we do not take into account additional relativistic c.m. corrections in Eq.~\eqref{eq:HamMassCorr}, for example $\hat{p}^4/c^2$ or $g\hat{p}\hat{z}\hat{p}/c^2$.
While such contributions are of order $c^{-2}$, they are independent of the mass defect $\Delta m$.
Consequently, they cancel in the differential phase between two internal states with $m_a$ and $m_b$.
Similarly, we neglect effects from a modification of the wave vector due to the propagation of the light in gravity.
Such effects are to lowest order in the perturbation also independent of $\Delta m$.
In analogy to Eq.~\eqref{eq:PhaseIdealSingleClock1}, we consider $\Delta m\,\beta_j$ as a higher-order term that has to be neglected in our treatment.

Moreover, the term $\Delta m c^2$ is much larger than the other terms in $\hat{\mathcal{H}}_j^{(\sigma)}$ and constitutes the dominant contribution to proper time $\tau$ along a given trajectory, as it measures the laboratory time $t$.
Although being dominant, this contribution does not modify the trajectories and is therefore exactly accounted for in the action.
In addition, it is common to both branches and cancels in the proper-time difference between them.
Because of the finite speed of light, each branch is addressed by the interacting light field at slightly different times. 
Thus, one branch accumulates an additional phase $(m\pm\Delta m/2)c^2\delta t/\hbar$ in a specific time interval $\delta t$.
However, between separation and detection both interferometer branches experience the same amount of laboratory time and therefore this phase cancels.
A similar argument applies not only to atom interferometers but also to the clock geometries considered in the following where the two clocks do not meet at the final pulse.
Next-order effects from the finite speed of light emerge from the remaining part of the unperturbed Hamiltonian from Eq.~\eqref{eq:HamMassCorr}. 
For example, for atom interferometers, these terms result in phases of the form $m\delta z g T/\hbar$, where $\delta z$ is the spatial separation of the branches.
Therefore, finite speed-of-light effects appear as $m\delta z g (T+\delta T)/\hbar$.
Since such effects are common to both internal states, they cancel in the differential phase between $m_a$ and $m_b$.
Thus, the remaining contributions arise from the perturbation Hamiltonian described by Eq.~\eqref{eq:PerturPot} and are proportional to $\Delta m$.
Moreover, the time difference $\delta T$ can be estimated as $\delta z/c$.
Consequently, these effects are at least of order $\Delta m/c$ in the differential phase and lie beyond our treatment~\cite{Loriani2019,Ufrecht2020}.

\subsection{Classical UGR tests by clocks with quantized motion}
\label{sec.UGR_with_atomic_clocks}
\begin{figure*}
	\centering
	\includegraphics[width=1\textwidth]{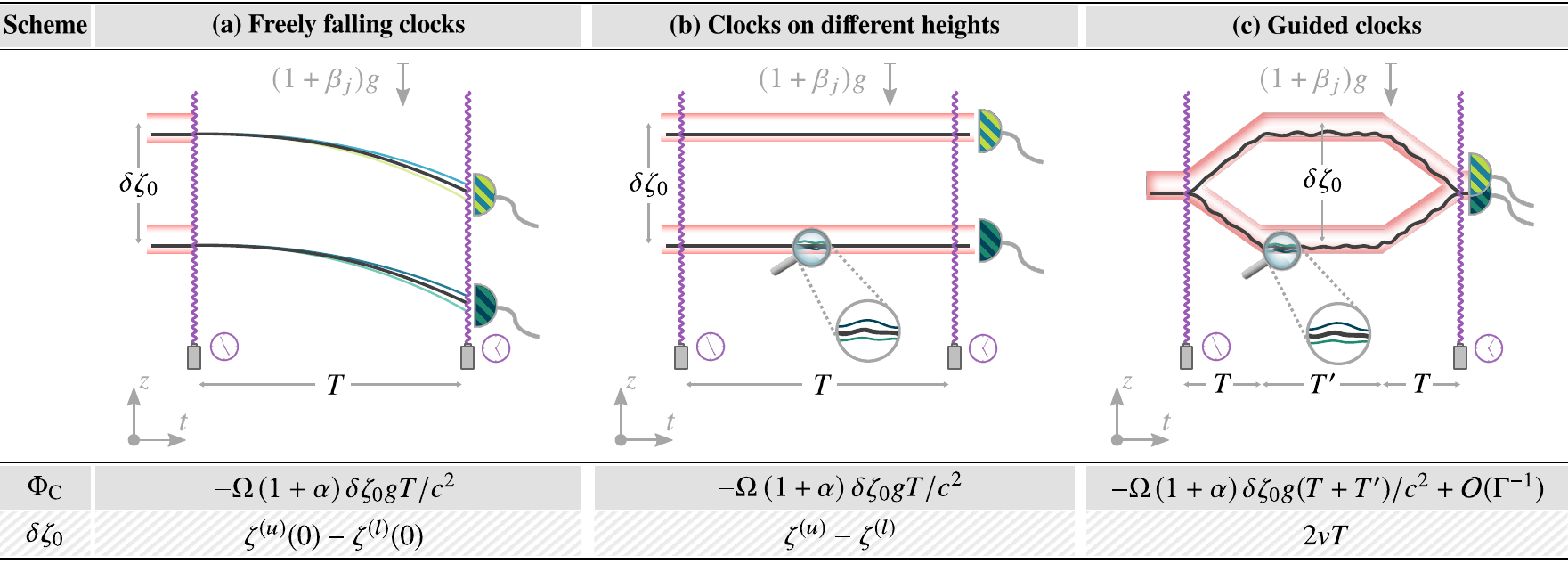}
	\caption{Comparison of two clocks moving along different configurations in spacetime:
	two freely falling atoms (a); two clocks trapped at different heights (b); as well as two clocks starting at the same position, guided by two trapping potentials to different heights, and back to the initial height (c).
	The center of each trap is bordered by red potential barriers, while the unperturbed trajectories $z(t)$ of the atoms are represented by black lines.
	The initialization and readout of the clocks is marked by purple pulses.
	To show that the detectors project on the internal state of the atom, each detector is colored in blue and green, representing both internal states.
	The two trajectories caused by a perturbation through the dilaton field with an effective acceleration $(1+\beta_j)g$ are also drawn with these colors.
	The differential phases $\Phi_\text{C}$ listed for each geometry in the table all show a sensitivity to UGR tests for a height difference of $\delta \zeta_0$ between the positions of the trap centers.
	}
\label{fig:DifferentClockGeom}
\end{figure*}
A single atomic clock, for example operated as a $\pi/2$-$\pi/2$ Ramsey sequence, relies on the overlap of the c.m. wave packets associated with two different internal states $\ket{a}$ and $\ket{b}$ on one branch $\sigma$ and measures the phase $\varphi^{(\sigma)}$.
In the following, we omit the superscript $\sigma$ when we discuss only a single atomic clock.
To describe a clock, we model the first $\pi/2$ pulse by assuming an initial internal superposition $\left(\ket{a}+\ket{b}\right)/2$.
Each of these two internal states $\ket{j}$ evolves according to its respective Hamiltonian given in Eq.~\eqref{eq:HamMassCorr} and gives rise to the effective operators $\hat{U}^{}_1$ and $\hat{U}^{}_2$.
Hence, in the case of atomic clocks, these two operators describe the c.m. motion of the atom in the ground state and in the excited state, respectively.
Furthermore, we neglect recoil effects and choose a vanishing force $F=0$.
We model the final $\pi/2$ pulse mixing the two internal sates and the readout through a projection onto the superposition $\left(\ket{a}+\ket{b}\right)/2$~(see Appendix~\ref{sec:interference_signal}).

Hence, we rely on the difference $\mathcal{H}_\text{diff}=\mathcal{H}_b-\mathcal{H}_a$, evaluated along the unperturbed trajectory $z(t)$.
Consequently, we find the phase
\begin{align}
    \label{eq:SinglePHaseOneClock}
    \varphi=&\!-\!\frac{1}{\hbar}\int\!\!\!\text{d}t\,\Bigg\lbrace \!\Delta m\bigg[c^2-\frac{\dot{z}^2}{2}+g z\bigg]\!\!+m\Delta\beta g z+\!\frac{m\Delta\Gamma^2}{2}(z-\zeta)^2\Bigg\rbrace \nonumber \\
    &+\varphi_\text{WP}
\end{align}
measured by one clock.
Here we omit the phase $\varphi_0$, because the unperturbed clock (with respect to the mean mass $m$) closes trivially and  $\varphi_0$ contains only laser phases.
These phases vanish in differential measurements if the same laser is used for both clocks or if the lasers for each clock are phase locked.
Because the c.m. wave packets associated with the two internal states may have obtained a different shape upon propagation, wave-packet effects $\varphi_\text{WP}$ have to be taken into account as well.

Since classical UGR tests are based on the comparison of two clocks, we introduce the differential phase 
\begin{equation}
    \label{eq:DiffPhaseClocks}
    \Phi_\text{C}=\varphi^{(u)}-\varphi^{(l)}
\end{equation}
between two clocks on different branches as a generalization of Eq.~\eqref{eq:PhaseIdealClockOnTwoHeights}.
We perform a partial integration of Eq.~\eqref{eq:SinglePHaseOneClock}, which corresponds to the application of the virial theorem, and utilize the classical, unperturbed equation of motion~(see Appendix~\ref{sec:ClassEqMot}).
Introducing the mean position $\bar z=\left(z^{(u)}+z^{(l)}\right)/2$ and the difference $\delta z=z^{(u)}-z^{(l)}$, which are defined analogously for $\zeta$, we find the differential phase
\begin{align}
   \label{eq:DiffPhaseClocksPartInt}
   \Phi_\text{C}=-\frac{1}{\hbar}\Big[&\left.-\Delta m\dot{\bar z}\delta z\right|+\int\!\!\text{d}t\,\Delta m\left\lbrace\Gamma^2(\bar\zeta-\bar z)+\alpha g\right\rbrace\delta z \nonumber \\
   &+\int\!\!\text{d}t\,m\Delta\Gamma^2(\bar\zeta-\bar z)(\delta\zeta-\delta z)\Big].
\end{align}
As shown in Appendix~\ref{sec:WPEffects}, the wave-packet contribution $\varphi_\text{WP}$ is branch independent and cancels in the differential phase.
With this result, we discuss the different situations shown in Fig.~\ref{fig:DifferentClockGeom}, including a quantized c.m. motion of the clocks.

The configuration in Fig.~\ref{fig:DifferentClockGeom}\hyperref[fig:DifferentClockGeom]{(a)} consists of two initially stationary atoms at different heights trapped until their initialization at time $t=0$.
After their release, i.\,e., $\Gamma=0$, the atoms fall freely in a dilaton gravitational field.
Therefore, we find for the unperturbed trajectory that $\dot{\bar z}(T)-\dot{\bar z}(0)=-gT$ and we observe that $\delta z=\delta\zeta_0$ is constant.
Consequently, only the first term and the integral over $\Delta m\alpha g \delta z$ in Eq.~\eqref{eq:DiffPhaseClocksPartInt} contribute to the differential phase and lead to a UGR test.
Indeed, two atomic clocks in free fall are suitable to measure UGR violations.
The key step is that the clocks are initialized in traps at different heights, an assumption that we drop in the context of atom interferometers.
As a direct consequence, conventional Bragg-type atom interferometers will turn out to be insensitive to UGR violations.

If the atoms are trapped, i.e., $\Gamma\neq0$ for the whole interference experiment, we find the difference between the mean trajectory and mean trap position from a general perturbative treatment in Appendix~\ref{sec:ClassEqMot}.
Through this procedure, we obtain $\bar z-\bar\zeta=-g/\Gamma^2+\mathcal{O}(\Gamma^{-3})$, which corresponds to order $\Gamma^{-2}$ to the gravitational sag.
The result is valid as long as $\ddot{\bar\zeta}=0$, i.\,e., if the two traps are accelerated in an antisymmetric fashion.
A residual mean acceleration $\ddot{\bar\zeta}$ would lead to an accelerational redshift instead of a gravitational one~\cite{Okolow2020}.
Similarly, the height difference between the atomic branches is $\delta z=\delta\zeta+\mathcal{O}(\Gamma^{-2})$. 
Hence, apart from corrections of order $\Gamma^{-2}$, it corresponds to the height difference between the positions of the trap centers.
Thus, we obtain for the differential phase
\begin{align}
   \label{eq:DiffPhaseClocksPartIntAndInserted}
   \Phi_\text{C}=-\frac{1}{\hbar}\Big[\left.\!-\Delta m\dot{\bar\zeta}\delta\zeta\right|+\!\int\!\!\text{d}t\,\left\lbrace\Delta m\left(1+\alpha\right) \delta\zeta g\right\rbrace+\mathcal{O}(\Gamma^{-1})\Big],
\end{align}
where the first term arising from the boundary conditions is independent of $g$.
For the derivation we assumed that different trap frequencies can be treated perturbatively, which is valid for $\Delta\Gamma\ll\sqrt{\Gamma/T}$, as shown in Appendix~\ref{sec:WPEffects}.
Additional corrections from different trap frequencies are of order $\Delta\Gamma^2/\Gamma^4$.

Figure~\ref{fig:DifferentClockGeom}\hyperref[fig:DifferentClockGeom]{(b)} shows a situation where the atoms evolve for time $T$ in two traps at different but constant heights separated by $\delta\zeta_0$. 
Hence, the time derivative of $\bar\zeta$ and by that the first term in Eq.~\eqref{eq:DiffPhaseClocksPartIntAndInserted} vanishes.
Thus, the expression in the integral leads to the UGR sensitivity.

If both clocks meet at the beginning and the end of the sequence, the first term in Eq.~\eqref{eq:DiffPhaseClocksPartIntAndInserted} also vanishes.
Hence, the differential phase $\Phi_\text{C}=-\Omega\left(1+\alpha\right)g\!\int\!\!\text{d}t\,\delta\zeta/c^2+\mathcal{O}(\Gamma^{-1})$ is solely characterized by the time-dependent difference of the trap centers.
One example for this class of geometries is the configuration from Fig.~\ref{fig:DifferentClockGeom}\hyperref[fig:DifferentClockGeom]{(c)}, which resembles a guided atom interferometer.
In this geometry, one trap moves an atom up for some time $T$ with velocity $v$, while another trap moves a different atom down with $-v$.
Afterwards, both traps are held at constant heights for an interval $T^\prime$, before the velocities are inverted and the atoms are combined at $t=2T+T'$, which yields $\int\!\!\text{d}t\,\delta\zeta=2vT (T+ T^\prime)$.
Hence, $2vT=\delta\zeta_0$ represents the height difference during the central time segment.
For this example, we did not accelerate the traps gradually but with instantaneous velocity kicks $\pm v$. 
Still, similar relations $\bar z-\bar\zeta=-g/\Gamma^2$ as well as $\delta z=\delta\zeta+\mathcal{O}(\Gamma^{-1})$ are valid for these instantaneous kicks due the antisymmetry of the velocity transfer.
In the limit $\Gamma\rightarrow\infty$ all contributions in $\Phi_\text{C}$ distinguishing $\bar z$ and $\bar\zeta$, as well as $\delta z$ and $\delta\zeta$ vanish. 
Since an infinitely steep trap enforces the colocalization of the actual position $z(t)$ with the trap center $\zeta(t)$, we recover idealized clocks in this limit. 
For the scheme of two clocks in free fall, no trap enforces the colocalization after the atom's release at $t=0$, but the wave-packet effects cancel in $\Phi_\text{C}$ since they are branch independent.

The table of Fig.~\ref{fig:DifferentClockGeom} summarizes our results and gives the differential phase as well as the definition of the height difference.
All three situations test UGR \emph{exactly} like idealized clocks, that is, with the internal frequency $\Omega$ and the factor $(1+\alpha)$.
Any experiment giving rise to a phase of the form $\pm\Omega\left(1+\alpha\right)\delta\zeta_0 g T/c^2$ could have also been performed with two (static) atomic clocks with frequency $\Omega$ and separation $\delta\zeta_0$, operating for a time interval $T$.
Hence, such a phase defines \emph{the} gold standard for tests of UGR.
This UGR sensitivity relies on the application of a quadratic potential.
The resulting term $\Delta m\Gamma^2\bar z\delta z$ under the integral in Eq.~\eqref{eq:DiffPhaseClocksPartInt} persists also after a transformation into a freely falling system.
For freely falling clocks this role is played by the boundary term $\left.-\Delta m\dot{\bar z}\delta z\right|$ in Eq.~\eqref{eq:DiffPhaseClocksPartInt}. 
It is only nonvanishing because of the application of a quadratic potential at $t=0$ to bring the atoms at different heights.

The expressions for the differential phases in Fig.~\ref{fig:DifferentClockGeom} have been derived under the assumption that both internal states are initially at rest and colocalized in the unperturbed potential minimum.
Dropping this assumption, one is limited by additional requirements on the maximum amplitude of the classical oscillation as outlined in Ref.~\cite{Moller1956}.
However, our treatment goes beyond these classical considerations by including quantum-mechanical wave-packet effects for atomic clocks, which are described in more detail in Appendix~\ref{sec:WPEffects}.
In this article, we consider only atomic clock geometries without internal state transfer during the interrogation time.  
However, modeling the internal state transfer by a time-dependent mass, the treatment can be easily generalized.
As we will discuss in Sec.~\ref{sec.UGR_variable_mass}, such transitions are a possible key to obtain a sensitivity to UGR violations for atom interferometers with linear interaction potentials.

\subsection{UGR tests with quantum clock interferometry}
\label{sec.UGR_with_QCI}
Next we study atom interferometers where the two branches are not associated with two independent quantum systems but with a single one in spatial superposition.
Our aim is to examine which atom-interferometric configurations exhibit the same UGR sensitivity as atomic clocks.
Therefore, we first consider Bragg-type atom interferometers operated with a single internal state during the whole sequence, without any internal transitions.
The exit port of Bragg-type atom interferometers is defined by a specific momentum and its readout is performed, for example, through absorption imaging in the far field.
The interference pattern is determined by the overlap of the two c.m. wave packets that propagated along two different branches $\sigma = u, l$, measuring the phase $\varphi_j$.
Thus, the operators $\hat{U}^{}_1$ and $\hat{U}^{}_2$ are associated with the effective evolution of the c.m. wave packet along the branches.
In the following, we omit the index $j$ when we discuss an atom interferometer operated with only one internal state.
Because only one internal state is populated during the whole sequence, Bragg-type atom interferometers cannot test UGR like atomic clocks that necessarily involve two internal states.
As a generalization, we introduce quantum clock interferometry where an interferometer sequence is performed simultaneously for a superposition of internal states.
We furthermore investigate if such schemes are UGR sensitive.
Finally, we briefly discuss the connection of guided atom interferometers to atomic clocks guided in traps.

\subsubsection{Bragg-type atom interferometers}
In contrast to atomic clocks, atoms in Bragg-type atom interferometers are in free fall or manipulated by a linear, state-independent interaction potential, e.g., light pulses, instead of a harmonic trap.
Hence, we set $\Gamma=0$.
The light pulses introducing momentum kicks or other constant forces are encoded in a branch-dependent force $F^{(\sigma)}$.
Thus, we find the difference $\mathcal{H}_{\text{diff}}=\mathcal{H}^{(u)}-\mathcal{H}^{(l)}$ between the perturbation evaluated along the upper and lower unperturbed trajectory.
Thus, we obtain from Eq.~\eqref{eq:GeneralPhasePerturb} the phase 
\begin{align}
\label{PhaseAIBeginn}
    \varphi=&\varphi_0-\frac{1}{\hbar}\int\!\text{d}t\,\left\lbrace\lambda_j\frac{\Delta m }{2}\left[-\dot{\bar z}\delta\dot z+g\delta z\right]+m\beta_j g\delta z\right\rbrace \nonumber \\
    &+\varphi_{\text{WP}}
\end{align}
for a single internal state $\ket{j}$, expressed through the mean position $\bar z$ and the difference $\delta z$ between the branches.
An imperfect overlap between the two c.m. wave packets leads to the phase $\varphi_{\text{WP}}$.
However, since we consider unperturbed atom interferometers that are closed in phase space, no wave-packet effects arise and $\varphi_{\text{WP}}$ cancels~(see Appendix~\ref{sec:WPEffects}).

We cast Eq.~\eqref{PhaseAIBeginn} into a more compact form by partial integration and by using the classical, unperturbed equation of motion~(see~Appendix~\ref{sec:ClassEqMot}), which corresponds to the application of the virial theorem. 
Consequently, closed schemes measure the phase 
\begin{align}
   \label{eq:PhaseDifferenceBraggAI}
   \varphi=-\frac{1}{\hbar}\int\!\text{d}t\,\left\lbrace\lambda_j\frac{\Delta m }{2m}\bar F+ m\beta_j g\right\rbrace\delta z+\varphi_0,
\end{align}
which differs significantly from the UGR test through atomic clocks from Eq.~\eqref{eq:PhaseIdealClockOnTwoHeights}.
It implies that (closed) Bragg-type atom interferometers cannot provide such tests.
The integrand shows a term proportional to $\bar F$ that will be discussed later, and a term proportional to $\beta_j$.
In particular, we find for a Mach-Zehnder configuration the phase $\varphi=-k \left(1+\beta_j\right)g T^2$ between the two branches, where $T$ is the time between the pulses and $k$ the transferred momentum.
In contrast to the UGR violation with clocks, this phase depends on $\beta_j$ instead of $\alpha$ and does not include a reference energy of a second internal state.
As we discussed in Sec.~\ref{sec.Intr}, this result is already known from discussions of Mach-Zehnder interferometers~\cite{Wolf2011}, but Eq.~\eqref{eq:PhaseDifferenceBraggAI} generalizes it to all (closed) Bragg-type atom interferometers.
Moreover, the appearance of $\beta_j$ already highlights that such an interferometer can be used for tests of UFF, as the gravitational acceleration can in principle be compared to any other (even macroscopic) object.

\begin{figure*}
	\centering
	\includegraphics[width=1\textwidth]{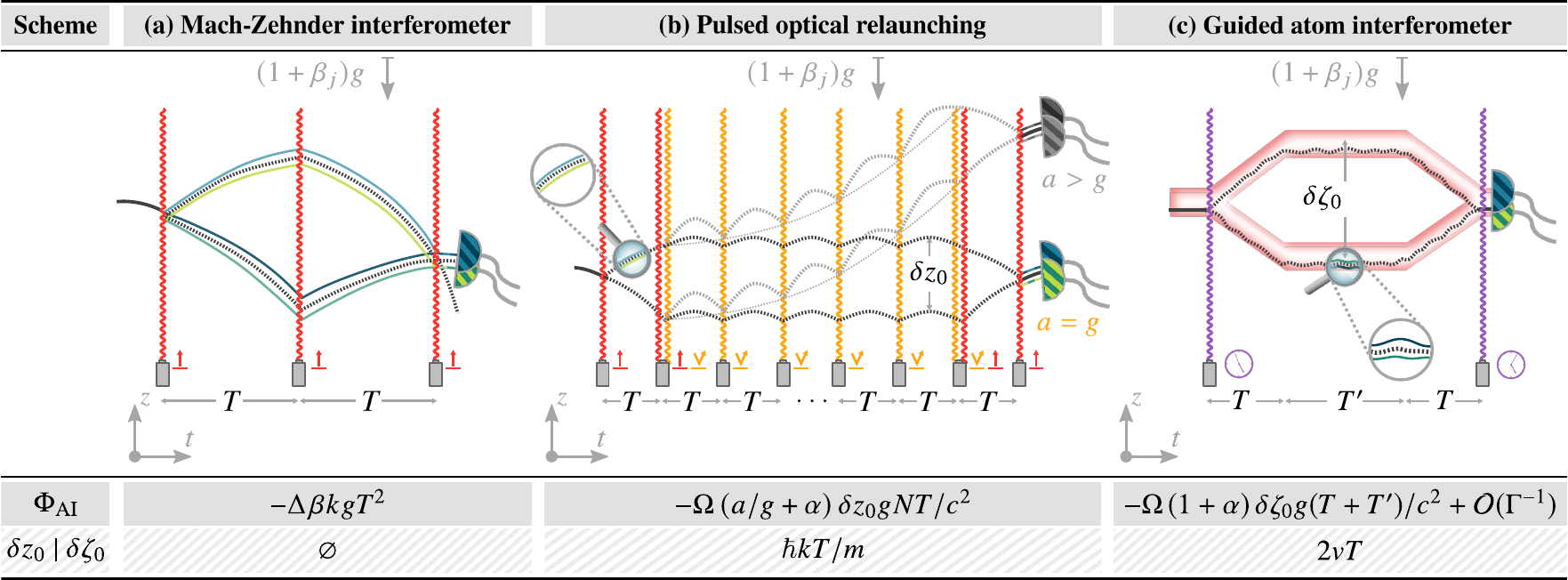}
	\caption{Spacetime diagrams of different quantum-clock atom interferometers:
	a conventional Mach-Zehnder interferometer (a); a scheme relying on pulsed optical relaunching (b); and an interferometer guided by two trapping potentials to different heights, and back to the initial height (c).
	The center of each trap is bordered by red potential barriers, while the unperturbed trajectories $z(t)$ of the atoms are represented by dashed black lines, highlighting the superposition between the arms of the interferometer.
	The diffracting Bragg pulses that act differently on each branch are denoted by red pulses.
	The yellow pulses are used for pulsed optical relaunching that is independent of the state and branch.
	One can measure the exit-port population for each internal state independently. 
	To indicate this measurement scheme, we introduced two separate detectors colored in blue and green.
	Accordingly, the trajectories of each internal state caused by a perturbation through the dilaton field with an effective acceleration $(1+\beta_j)g$ are also drawn with these colors.
	The guided atom interferometer in panel (c) takes the same form as for two guided clocks with recoilless initialization and readout pulses, with the only difference that the atom is in a superposition of the branches (indicated by the dashed unperturbed trajectory). 
	The differential phases $\Phi_\text{AI}$ are listed for each geometry in the table below.
	While the Mach-Zehnder interferometer in panel (a) is insensitive to the gravitational redshift, the pulsed optical relaunching in panel (b) mimics a UGR test for the special choice $a=g$, which corresponds to a vanishing effective acceleration and perfect levitation in the central segment.
	For other choices of $a$, the resulting effective acceleration leads on average to curved trajectories of the atoms, indicated by the grayed-out configuration. 
	In this case one observes an accelerational redshift.
	The table also includes the respective height differences $\delta z_0$ or $\delta \zeta_0$.
    }
\label{fig:DifferentAIGeom}
\end{figure*}
So far, $\Delta m$ appeared only as a technical quantity because there is no second internal state, but it is crucial for the concept of quantum clock interferometry:
instead of a single internal state $\ket{j}$, one can use a superposition $\left(\ket{a}+\ket{b}\right)/\sqrt{2}$ of internal states as input to a Bragg-type atom interferometer~\cite{Sinha2011}.
As a result, the interference signal is a superposition of two patterns, each associated with an internal state and the phase $\varphi_j$. 
A possible mass dependence of this phase introduces a beating of the interferometer signal~\cite{Zych2011,Loriani2019}, which effectively leads to a differential measurement of both phases.
Instead of measuring the differential phase through a beating, the phases of each internal state can be read out independently and subtracted.
From a fundamental perspective, superpositions of internal states are not necessary and the differential phase of two independent interferometers performed with different internal states leads to the same result.
However, superpositions can be beneficial to suppress common-mode effects.
In all these cases we infer the differential phase
\begin{equation}
    \label{eq:DiffPhaseAIs}
    \Phi_\text{AI}=\varphi_b-\varphi_a
\end{equation}
between two phases $\varphi_j$ of the individual states $\ket{j}$, where the unperturbed phase $\varphi_0$ cancels in such schemes.
Analogously to clocks, we find as a general result with the help of Eq.~\eqref{eq:PhaseDifferenceBraggAI} the differential phase
\begin{align}
   \label{eq:DiffPhaseAIsPartInt}
   \Phi_\text{AI}=-\frac{1}{\hbar}\int\!\text{d}t\,\left\lbrace\frac{\Delta m}{m}\bar F+ m\Delta\beta g\right\rbrace\delta z.
\end{align}
Here, $\bar F=\big(F^{(u)}+F^{(l)}\big)/2$ is the time-dependent mean force of the linear interaction potential.
It is independent of $g$ in contrast to the analogous term for clocks in Eq.~\eqref{eq:DiffPhaseClocksPartIntAndInserted} and does not change in the absence of gravity.
Since, for a linear gravitational potential, one can transform the unperturbed trajectories of both branches into the same freely falling frame, their difference $\delta z$ is solely determined by the momenta transferred by the interaction with the lasers and thus independent of gravity~\cite{Loriani2019}.
Relying on the relation $\alpha=m\Delta\beta/\Delta m$, one finds an integrand \smash{$\Omega(\bar F/m+\alpha g)\delta z/c^2$}, instead of $\Omega(1+\alpha)g\delta z/c^2$, which shows that such schemes measure in general the \emph{accelerational} redshift~\cite{Okolow2020} caused by an acceleration $\bar F/m$ instead of $g$.
However, it is possible to mimic the gravitational redshift by fixing $\bar F/m=g$ or $\int\!\text{d}t\,\bar F\delta z/m=g\int\!\text{d}t\,\delta z$, so that these schemes measure the same differential phase as two atomic clocks, even though the acceleration does not necessarily have to be of gravitational origin.
Note that choosing the mean force $\bar F$ implies that both internal states will have slightly different velocities because the mass $m_j$ and the dilaton violation parameter $\beta_j$ depend on the internal state. 
These different velocities lead to a divergence of their true, perturbed trajectories.
Yet, for appropriate time scales of the experiment, these deviations are included by our perturbative formalism and are treated consistently.
Equally, schemes that rely on Bragg-type specular mirrors to manipulate the atoms \cite{Giese2019,DiPumpo2020} also belong to this category of geometries for properly fixed mirror positions.
Below, we give an example for a modified Bragg-type geometry that resembles a UGR test.

Alternatively, if $\int\!\text{d}t\,\bar F\delta z/m$ is different from $g\int\!\text{d}t\,\delta z$, Bragg-type atom interferometers can be interpreted as UFF tests.
The prime example is the Mach-Zehnder scheme shown in Fig.~\ref{fig:DifferentAIGeom}\hyperref[fig:DifferentAIGeom]{(a)}.
Atom-interferometric UFF tests for this type of interferometer are often performed with different species~\cite{Schlippert2014}, so that there is no need for relativistic mass corrections in the description.
In such cases, the considered violation models couple directly to the rest mass and predict different gravitational accelerations for different rest masses.
However, recent works show that UFF tests are also possible for different internal states~\cite{Zhang2018} through $\Delta\beta$ in Eq.~\eqref{eq:DiffPhaseAIsPartInt}.
For the Mach-Zehnder geometry shown in Fig.~\ref{fig:DifferentAIGeom}\hyperref[fig:DifferentAIGeom]{(a)}, we find that $\int\!\text{d}t\,\bar F\delta z=0$ and $\Phi_\text{AI}=-\Delta\beta k g T^2$.
An analogous expression describes UFF tests of different atomic species or isotopes.
If performed with two different internal states, the differential phase can be connected by $\Delta \beta = \alpha \Delta m /m$ to the parametrization $\alpha$ used for UGR violations~\footnote{Note that the ratio $\Delta m / m \ll 1$ corresponds to the ratio of transition frequency to mean Compton frequency.}.
However, this measurement still represents a null test, so that this result is different to violations detected by atomic clocks, which measure modifications to proper time by a factor $(1  + \alpha)$, as discussed in Eq.~\eqref{eq:PhaseIdealClockOnTwoHeights}.

In order to modify this Mach-Zehnder scheme so that it becomes comparable to the clock geometry of Fig.~\ref{fig:DifferentClockGeom}\hyperref[fig:DifferentClockGeom]{(c)}, we study configurations where the atoms fall in parallel during the central time segment. 
In contrast to conventional Ramsey-Bord\'{e} configurations, both arms are relaunched~\cite{Hughes2009,Abend2016} in parallel through light pulses.
We explicitly model the light pulses by a branch-dependent but state-independent (magic Bragg~\cite{Katori2003}) effective potential~\cite{Loriani2019}. 
Thus, we find the mean force $\bar F=\hbar\sum_{\ell}\bar k_\ell\delta\left(t-t_\ell\right)$ with $\bar k_\ell=(k^{(u)}+k^{(l)})/2$, where $k^{(\sigma)}_\ell$ denotes the effective wave vector. 
Furthermore, $t_\ell$ is the time of the $\ell$-th laser pulse acting on branch $\sigma$.
Additionally, we also include $N$ relaunch pulses transferring the momentum $\hbar \kappa_\ell$ after equidistant time intervals $T$ to both branches.
We express this transfer $\kappa_\ell=ma T/\hbar$ through an effective acceleration $a$ and find the grayed-out configuration in Fig.~\ref{fig:DifferentAIGeom}\hyperref[fig:DifferentAIGeom]{(b)}.
This acceleration $a$ can be tuned by adjusting the time interval $T$.
The first two pulses as well as the last two pulses are also separated by the interval $T$. 
They are associated with conventional branch-dependent momentum kicks that lead to the separation $\delta z_0=\hbar k T/m$ and the closing of the interferometer.
After this separation, we apply $N$ relaunch pulses to these closed Ramsey-Bord\'{e}-like configurations that add to the mean force $\bar F$.
In general, the resulting differential phase takes the form \smash{$\Phi_\text{AI}=-\Omega\left(a/g+\alpha\right)\delta z_0 g N T/c^2$}, which measures the accelerational redshift caused by $a$. 
For $a=0$ and $N=1$, this geometry recovers the Mach-Zehnder scheme shown in Fig.~\ref{fig:DifferentAIGeom}\hyperref[fig:DifferentAIGeom]{(a)}.

For arbitrary accelerations $a$, we observe from the grayed-out configuration in Fig.~\ref{fig:DifferentAIGeom}\hyperref[fig:DifferentAIGeom]{(b)} that the atoms still experience an effective acceleration in the central segment.
So in order to resemble the clock geometry from Fig.~\ref{fig:DifferentClockGeom}\hyperref[fig:DifferentClockGeom]{(c)} as closely as possible, one has to ensure that the atoms are ideally levitated. 
In this context, ideal levitation means that effectively no acceleration is imparted to the atoms during the relaunch sequence, i.\,e., between the second beam splitter and the last relaunch pulse.
Consequently, we have to fix $a=g$ by tuning the time $T$ appropriately, which corresponds to straight lines for the time-averaged, unperturbed trajectories of the atoms.
This situation simulates two atoms in stationary harmonic traps between $t=T$ and $t=NT$, so that the differential phase of two atomic clocks is reconstructed.
In this case, the differential phase $\Phi_\text{AI}=-\Omega\left(1+\alpha\right)\delta z_0 g N T/c^2$ from the colored scheme in Fig.~\ref{fig:DifferentAIGeom}\hyperref[fig:DifferentAIGeom]{(b)} mimics UGR tests with atomic clocks.
Although both internal states possess a different mass $m_j$ and a different gravitational acceleration $(1+\beta_j)g$, we can still choose the light pulses to transfer the same momentum to both internal states, as long as the requirements for our perturbative treatment from Eq.~\eqref{eq:SinglePHaseOneClock} are fulfilled.
In particular, these requirements include a time scale for the interrogation time of the interferometer, and are amply fulfilled in state-of-the-art experiments.

In contrast, if the requirements for our perturbative treatment are not fulfilled, for example by a very long interrogation time, using state-independent light pulses and forces will lead to significantly diverging trajectories of both internal states.
This result is due to the different masses $m_j$ and different gravitational accelerations $(1+\beta_j)g$.
To resemble situations where the atoms are trapped in a harmonic potential in such extreme situations, $a$ has to be chosen state dependently via $m_j$ and $\beta_j$.
Since $\beta_j$ is not known prior to the experiment, this situation reveals the difference between Bragg-type atom interferometers and atomic clocks, because for the latter, adjusting the trapping potential according to $\beta_j$ is not necessary. While this discussion highlights the fundamental difference between atomic clocks and atom interferometers in the context of UGR tests, such limitations of the perturbative description are of no practical concern.

Recently established experimental techniques~\cite{Xu2019} in cavity-based compact devices allow levitating atoms in a spatial superposition for unprecedented times.
Although the Bloch oscillations used in this particular experiment differ from our pulsed scheme, these results pave the way towards mimicked UGR tests at hitherto inaccessible ranges of parameters. 
As such they emphasize the potential and importance of levitated schemes.
Together with an internal superposition such setups measure differential phases similar to the one discussed in Fig.~\ref{fig:DifferentAIGeom}\hyperref[fig:DifferentAIGeom]{(b)}.
Thus, the concepts established in this article can play a major role in future UGR tests with atom interferometry, pushing the sensitivity towards current limits set by atomic clocks. 
We underline this statement by a rough estimate in the discussion of Sec.~\ref{sec.Discussion}.
Conventional light-pulse atom interferometers without relaunch pulses are insensitive to UGR violations~\cite{Loriani2019}, as discussed before. 
Conversely, our results show that, in contrast to conventional setups, levitating pulses can indeed mimic UGR tests.
They therefore highlight the importance of the exact pulse arrangement and momentum transfer to test UGR with atom interferometers.

\subsubsection{UGR sensitivity through state-dependent, linear potentials}
If the diffraction mechanism for the matter wave were to allow for a velocity instead of a momentum transfer~\cite{Paige2019}, or more generally a state-dependent, linear potential~\cite{Amit2019}, one could indeed test UGR. 
In this situation a term $-\lambda_j\Delta m \mathcal{F}^{(\sigma)}\hat z/(2m)$ adds to the perturbation Hamiltonian described by Eq.~\eqref{eq:PerturPot}, with a modified branch-dependent, linear force $\mathcal{F}^{(\sigma)}$.
Consequently, for closed schemes and state-independent, unperturbed trajectories, Eq.~\eqref{eq:DiffPhaseAIsPartInt} takes the form $\Phi_\text{AI}=-\int\!\text{d}t\,\left\lbrace-\Delta m\bar z\delta \mathcal{F}/m+ m\Delta\beta g\delta z\right\rbrace/\hbar$, with $\delta \mathcal{F}=\mathcal{F}^{(u)}-\mathcal{F}^{(l)}$. 
Since the first term within the brackets depends on gravity through $\bar z$, one finds a UGR sensitivity akin to atomic clocks if $-\int\!\text{d}t\,\bar z\delta\mathcal{F}/m=g\int\!\text{d}t\,\delta z$.
We conclude that the crucial difference between Bragg-type atom interferometers and this mechanism is the state dependency of the linear potential. The resulting momentum change in these diffraction processes depends on the internal state and leads already in the Newtonian case to a term including $g$ times $\Delta m$. 
This coupling also arises in a freely falling reference frame.
However, all established diffraction mechanisms for light pulses so far rely on momentum rather than velocity transfers. 
Still, state-dependent linear potentials can indeed be realized by magnetic fields~\cite{Amit2019}, which may allow for UGR-sensitive configurations.

\subsubsection{Guided atom interferometers}
Instead of Bragg-type geometries with linear and state-independent potentials, atom interferometers can also be realized in fully guided schemes~\cite{Dumke2002,Berrada2013,Ryu2015,Navez2016}. 
In such guiding schemes, the laser does not induce momentum kicks by diffracting the atom from a light grating but merely traps the atom, for example, in a harmonic potential.
To model such guiding mechanisms that also act on superpositions of internal states, we apply a trapping potential similar to clocks.
Because of the harmonic potential, the Newtonian part of the differential phase $\Phi_\text{AI}$ ceases to be purely kinematic, in contrast to Bragg-type atom interferometers discussed before.
Moreover, both internal states are forced on one common trajectory $\zeta(t)$ like two independent atomic clocks.
Consequently, we find for closed schemes, i.\,e., for deep traps, that all wave-packet effects cancel in the differential phase.
As shown in Fig.~\ref{fig:DifferentAIGeom}\hyperref[fig:DifferentAIGeom]{(c)}, the phase measured by such a guided quantum-clock interferometer coincides with that obtained for clocks, and we find $\Phi_\text{AI}=-\Omega\left(1+\alpha\right)g\!\int\!\!\text{d}t\,\delta\zeta/c^2$.
This result demonstrates that guided quantum-clock interferometers are sensitive to UGR violations.
Indeed, it is not necessary to adjust the trapping potential to $\beta_j$ in this case.

Time-dependent double well potentials~\cite{Schumm2005,Berrada2013}, deep counterpropagating twin lattices~\cite{Pagel2019,Gebbe2021}, or spin-dependent optical lattices~\cite{Ke2018} constitute promising candidates for beam splitting mechanisms in such a scheme.
Moreover, semiguided geometries have been explored~\cite{Zhang2016,Xu2019}, where the separation is realized by momentum kicks, before the atom is trapped by a lattice. 

\section{UGR tests caused by internal transitions}
\label{sec.UGR_variable_mass}
So far, we have discussed the UGR sensitivity for atom interferometers in linear and in harmonic potentials.
However, we have not yet considered the possibility of driving internal transitions during the sequence.
As shown in Refs.~\cite{Roura2020,Ufrecht2020,Roura2020b}, this additional ingredient can lead to a UGR sensitivity in light-pulse atom interferometers.
Thus, we investigate this situation in more detail in the following.
For that, we consider single interferometric phases as well as differential phases together with mass changes caused by such transitions.
\begin{figure*}
	\centering
	\includegraphics[width=1.0\textwidth]{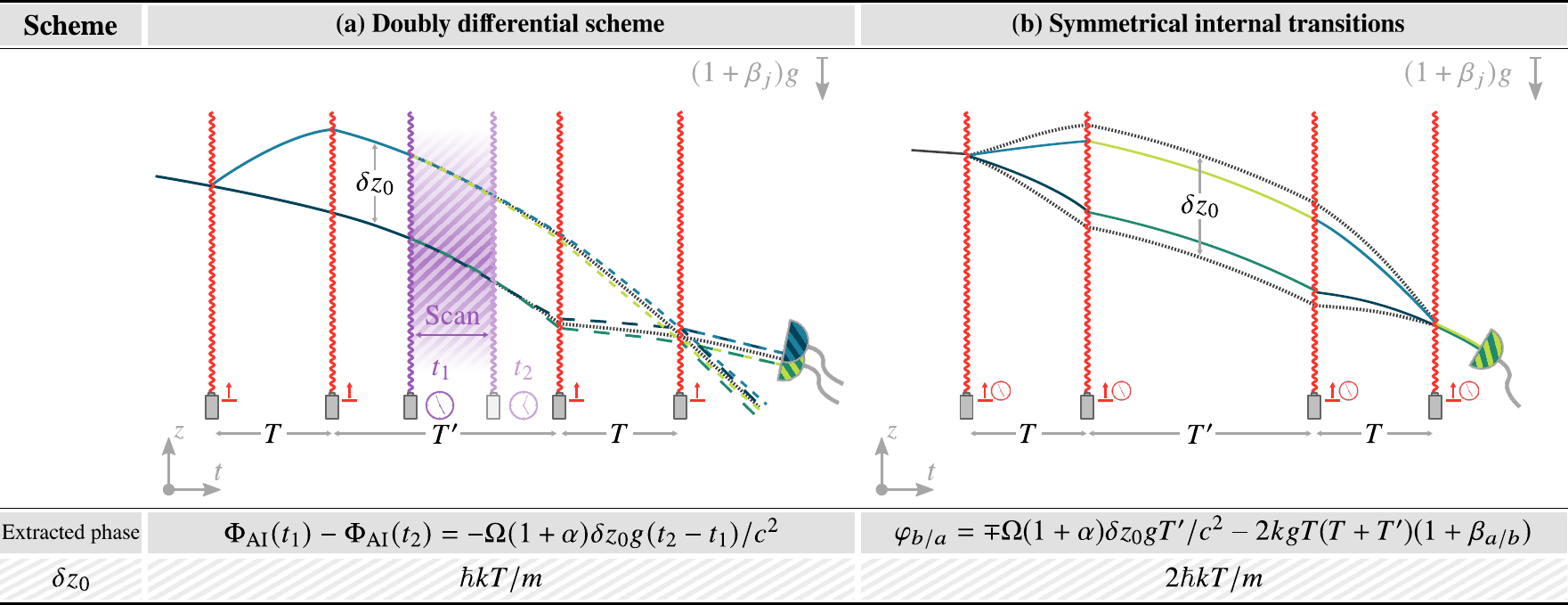}
	\caption{Spacetime diagrams of two UGR-sensitive atom interferometer geometries with internal transitions:
	a Ramsey-Bord\'{e}-like geometry used in a doubly differential scheme (a) and a scheme relying on successive symmetrical transitions of the internal state (b).
	The unperturbed trajectories $z(t)$ of the atoms are represented by dashed black lines, highlighting their superposition.
	To indicate that the exit-port population  for such interferometers can be measured independently for each internal state, we introduce two separate detectors that are colored blue and green.
	Accordingly, the trajectories of each internal state caused by a perturbation through the dilaton field with an effective acceleration $(1+\beta_j)g$ are also drawn in these colors.
	The redshift sensitivity of the Ramsey-Bord\'{e}-like geometry in panel (a) relies on a doubly differential scheme where the first two Bragg pulses denoted by red lines open the interferometer.
	While the atom drops in superposition of a height difference $\delta z_0$, it is brought into an internal superposition shown by a purple pulse at a time $t_1$ initializing the clock.
	The differential phase $\Phi_\text{AI}(t_1)$ between the interference patterns for each individual state depends on the initialization time.
	By subtracting the differential phase of another initialization time $t_2$, one measures the gravitational redshift between the two realizations, marked by the purple shaded area.
	In contrast, the configuration of panel (b) relies on symmetrical internal transitions that are combined with momentum transfers, symbolized by pictograms next to the pulse.
	Similar to panel (a), the atom travels during a central time segment in a superposition of parallel dropping trajectories.
	The phase $\varphi_{b/a}$ differs for two different input states (only one situation is depicted) and is already sensitive to UGR violations, even though additional terms arise so that the interferometer has to be performed with a differential measurement.
	The table below the figures lists the respective phases as well as the height differences $\delta z_0$.}
\label{fig:DifferentAIGeom2}
\end{figure*}

\subsection{Interferometric phase with internal transitions}
Transitions that change the internal state directly imply that the mass varies during the interferometer sequence and becomes time dependent.
Similarly, the violation parameter of the dilaton field is linked to the internal state, resulting in a time dependent $\beta(t)$ alternating between $\beta_a$ and $\beta_b$.
As a consequence, the perturbation Hamiltonian from Eq.~\eqref{eq:PerturPot} for $\Delta\Gamma=0$ takes the form 
\begin{equation}
    \hat{\mathcal{H}}^{(\sigma)}=\lambda^{(\sigma)}(t)\frac{\Delta m }{2}\left[c^2-\frac{\hat{p}^2}{2 m^2}+g\hat z\right]+m\beta^{(\sigma)}(t)g\hat{z},
\end{equation}
where $\lambda^{(\sigma)}(t)$ encodes the current internal state and corresponds to either $1$ or $-1$.
Additionally, branch-dependent light pulses directly imply in general a branch-dependent mass.
Thus, $\lambda\rightarrow\lambda^{(\sigma)}$ and $\beta\rightarrow\beta^{(\sigma)}$ depend on the branch.
This generalized Hamiltonian allows not only state- and momentum-changing Raman-type pulses, but also includes the effect of arbitrary changes of the mass through recoilless transitions.
We now assume that the internal state and consequently the dilaton violation parameter are common to both branches during a specific time interval.
By partial integration and utilizing the classical, unperturbed equation of motion~(see Appendix~\ref{sec:ClassEqMot}) the generalized phase reads
\begin{align}
    \label{eq:PhaseClosedRamanOneMass2}
    \varphi=&\varphi_0-\frac{1}{\hbar}\int\!\text{d}t\,\left\lbrace\frac{\Delta m }{2}\left[\frac{\lambda(t)}{m}\bar F+\dot{\lambda}(t)\dot{\bar z}\right]+m\beta(t)g\right\rbrace\delta z  \nonumber \\
    &+\left.\lambda(t)\frac{\Delta m }{2\hbar}\dot{\bar z}\delta z\right|+\varphi_{\text{WP}}.
\end{align}
Again, terms originating from the boundary conditions and the wave-packet effects vanish for closed geometries~(see Appendix~\ref{sec:WPEffects}).
The difference $\delta z$ is independent of gravitational contributions.
However, the mean velocity $\dot{\bar z}$ contains the gravitational acceleration $g$ and the derivative $\dot{\lambda}$ describing internal transitions is only nonvanishing during the instantaneous pulses.
It is exactly this term that allows for tests of UGR through light-pulse schemes employing internal transitions with the help of the interferometric phase alone.
Similar to state-dependent linear potentials discussed before, it introduces momentum changes that depend on the internal state before the pulse.
But in contrast to the former, this state-dependent momentum now arises for a single internal state, as it can be changed during the sequence.

\subsection{UGR-sensitive light-pulse schemes}
The configuration shown in Fig.~\ref{fig:DifferentAIGeom2}\hyperref[fig:DifferentAIGeom2]{(a)} is based on a symmetric Ramsey-Bord\'{e} scheme. 
The additional recoil-free $\pi/2$ pulse at time $t_1$ generates simultaneously (in the laboratory frame) a superposition of internal states on each branch~\cite{Roura2020}.
This way, the state independency of the momentum transfer is bypassed through this intermediate clock initialization.

The population of each of the atom's internal states, within a specific range of the final momentum of the atom, can be detected independently of the other internal state in an exit port.
This measurement scheme leads to the phase $\varphi_{a\veryshortrightarrow a}(t_1)=\varphi_a(0,t_1)+\varphi_a(t_1,t_\text{f})$. 
Here $\varphi_j(t,t^\prime)$ is the interferometric phase accumulated in state $\ket{j}$ from $t$ to $t^\prime$, when the atom is initially in the ground state $\ket{a}$ and detected in the same state.
In contrast, $\varphi_{a\veryshortrightarrow b}(t_1)=\varphi_a(0,t_1)+\varphi_b(t_1,t_\text{f})$ is the phase when the atom is initially in the ground state $\ket{a}$, and changes into state $\ket{b}$ at time $t_1$.
This transition at $t_1$ is associated with the time derivative $\dot{\lambda}(t)$.
The atom is finally detected at time $t_\text{f}$ in the excited state $\ket{b}$.
The transition introduces a term involving $g$ times $\Delta m$.
Hence, the differential phase $\Phi_\text{AI}(t_1)=\varphi_{a\veryshortrightarrow b}-\varphi_{a\veryshortrightarrow a}=\varphi_b(t_1,t_\text{f})-\varphi_a(t_1,t_\text{f})$ is independent of the phase accumulated during the time interval from zero to $t_1$.
We now calculate the phase for an instantaneous internal transition at $t_1$, so that the time derivative $\dot \lambda = 2 \delta (t-t_1)$.
For a closed interferometer, the differential phase based on Eq.~\eqref{eq:PhaseClosedRamanOneMass2} takes the form 
\begin{equation}
    \Phi_\text{AI}(t_1)=-\frac{1}{\hbar}\left[\Delta m \dot{\Bar{z}}(t_1)\delta z(t_1)+\int\limits_{t_1}^{t_\text{f}}\!\text{d}t\,\left\lbrace\frac{\Delta m}{m}\bar F+m\Delta\beta g\right\rbrace\delta z\right],
\end{equation}
where the first term is essential for a UGR test.
In the following we consider a double-differential scheme where we compare the phase from two different initialization times $t_1$ and $t_2>t_1$. 
By that we are able to separate the time segment in which both branches fall in parallel and remove the term proportional to $\bar F$ under the integral.
Indeed, the difference of both differential phases $\Phi_\text{AI}(t_1)-\Phi_\text{AI}(t_2)=\varphi_b(t_1,t_2)-\varphi_a(t_1,t_2)$ results in a phase solely determined by the time segment $t_2-t_1$ where the two branches move in parallel like the clocks in Fig.~\ref{fig:DifferentClockGeom}\hyperref[fig:DifferentClockGeom]{(a)}.
This scheme leads to
\begin{equation}
    \label{eq:ResultDDiff}
    \Phi_\text{AI}(t_1)-\Phi_\text{AI}(t_2)=-\frac{1}{\hbar}\left[\Delta m\left.\dot{\Bar{z}}\right|_{t_2}^{t_1}\delta z_0 +\int\limits_{t_1}^{t_2}\!\text{d}t\, m \Delta \beta g\delta z_0 \right],
\end{equation}
which resembles two clocks falling equidistantly with $\dot{\Bar{z}}(t_1)-\dot{\Bar{z}}(t_2)=g(t_2-t_1)$ and with a separation $\delta z(t_1)=\delta z(t_2)=\delta z_0 =\hbar k T /m$. 
Since the remaining integration can be performed trivially, we obtain the differential phase \smash{$ \Phi_\text{AI}(t_1)-\Phi_\text{AI}(t_2)=-\Omega\left(1+\alpha\right)\delta z_0 g (t_2-t_1)/c^2$}.

The presented scheme~\cite{Roura2020} resembles two atomic clocks, but even internal transitions without such superpositions can be used for UGR tests.
In fact, the configuration shown in Fig.~\ref{fig:DifferentAIGeom2}\hyperref[fig:DifferentAIGeom2]{(b)} relies on simultaneously changing on both branches the momentum and internal state, as well as on a symmetric diffraction into two opposite directions~\cite{Ufrecht2020}.
In such a case, there is no superposition of internal states and the geometry closes.
Because of its symmetry, the first term under the integral of Eq.~\eqref{eq:PhaseClosedRamanOneMass2} proportional to $\bar F$, which is independent of gravity, vanishes.
Moreover, the atom can be initially either in the excited state $\ket{b}$ or in the ground state $\ket{a}$ (interchanging the sequence of internal states in the figure).
Accordingly, with the two instantaneous transitions at $t_1=T$ and $t_2=T'+T$ we find the phase
\begin{align}
    \varphi_{b/a}=\varphi_0-\frac{1}{\hbar}\left[\pm\Delta m\left.\dot{\Bar{z}}\right|_{t_2}^{t_1}\delta z_0+\int\!\!\text{d}t\,m\beta(t)g\delta z\right]
\end{align}
for the atom detected in the excited state $\ket{b}$ or in the ground state $\ket{a}$, with $\delta z_0=2\hbar kT/m$.
The UGR sensitivity enters the phase solely through $\dot{\bar z}\dot{\lambda}$ from Eq.~\eqref{eq:PhaseClosedRamanOneMass2} and arises for each detection individually.
Here, the term from the boundary conditions has the same form as the analogous one in Eq.~\eqref{eq:ResultDDiff}.
However, it emerges already in the interferometric phase through the internal transitions, rather than in a doubly differential scheme.
The violation parameter $\beta(t)$ is the same in the first and final time segment and differs in the central one.
Moreover, it can be shown that $\delta z=\delta z_0 t/T$ in the first segment, $\delta z=\delta z_0(1-t/T)$ in the final one, and $\delta z=\delta z_0$ in central segment, so that the remaining integral yields $\delta z_0(\beta_{a/b}T+\beta_{b/a}T')$.
Consequently, the phase
\begin{equation}
    \label{eq:PhaseDoubleRaman}
	 \varphi_{b/a}=\mp\Omega\left(1+\alpha\right)\delta z_0 g T'/c^2-2kgT\left(T+T'\right)\left(1\!+\!\beta_{a/b}\right)
\end{equation}
can be divided into a UGR (first term) and a UFF (second term) sensitive contribution.
The latter arises from the unperturbed phase $\varphi_0=-2kgT (T+T^\prime)$ together with the UFF violation $\beta_{a/b}$.
The second term of Eq.~\eqref{eq:PhaseDoubleRaman} does not depend on $\Delta \beta$ but solely on $\beta_{a/b}$.
Therefore, comparing it to the gravitational acceleration of any other (even macroscopic) object determined in an independent experiment leads in principle to UFF tests.

By inverting the role of the internal states, the differential phase $\varphi_b-\varphi_a=-2\Omega\left(1+\alpha\right)\delta z_0 g T'/c^2+2k\Delta\beta gT\left(T+T'\right)$ depends on the factor $(1+\alpha)$ that arises in UGR tests with atomic clocks, as well as on the parameter $\Delta \beta$ that occurs in tests of UFF for two different internal states.
We stress again that both violation parameters are connected by $\Delta m \alpha = m \Delta \beta $ in our model.
However, in analogy to Fig.~\ref{fig:DifferentAIGeom}\hyperref[fig:DifferentAIGeom]{(a)}, the second contribution still represents a null test, so that there is again a difference compared to violations detected by atomic clocks.
These clock schemes measure modifications to proper time by a factor $( 1 + \alpha )$, as discussed in Eq.~\eqref{eq:PhaseIdealClockOnTwoHeights}.
Additionally, this differential scheme suppresses additional effects, for example from the relativistic c.m. corrections $\hat{p}^4/c^2$ or $g\hat{p}\hat{z}\hat{p}/c^2$ that otherwise arise in Eq.~\eqref{eq:PhaseDoubleRaman}.
Because a light pulse does not induce a transition simultaneously on both branches but with a time delay $\delta t$ that depends on the distance between them, yet another differential phase $\Omega \delta t$ arises.
However, because at resonance the frequency of the light corresponds to $\Omega$ and is imprinted during the interaction onto each branch, an additional phase $-\Omega\delta t$ cancels this contribution.
Therefore, due to the state-inverted scheme with the same momentum transfer $\hbar k$, further finite speed-of-light effects are suppressed with $1/c$ compared to the phase of interest~\cite{Ufrecht2020}.
Further terms from the propagation of the light field in gravitation would contribute without a differential measurement.
However, these state-independent terms are already suppressed by ten orders of magnitude compared to the phase of interest and can thus be neglected.
The contributions proportional to $(1+\alpha)$ and $\Delta \beta$ can be distinguished in this setup through a variation of the central time interval $T'$. 
Possible diffraction mechanisms~\cite{Hartmann2020} that fulfill the requirements set by this scheme are discussed in Refs.~\cite{Leveque2009,Alden2014,Ahlers2016}. 
Transferring the concept of subsequent internal transitions without an internal superposition~\cite{Ufrecht2020} to the doubly differential scheme~\cite{Roura2020} leads to a configuration~\cite{Roura2020b} where internal transitions are performed independently of the diffracting pulse. 
Hence, no magic Bragg diffraction is necessary for the implementation.
Thus, this combination of both ideas offers a possibility to circumvent some of the issues caused by diffraction and vibration noise.
However, this modification can be readily treated as another special case of our general result from Eq.~\eqref{eq:PhaseClosedRamanOneMass2} and implies that the UGR sensitivity originates from the same mechanism discussed in Eq.~\eqref{eq:PhaseClosedRamanOneMass2}.
The scheme therefore relies on the same fundamental working principles, but introduces notable experimental advantages.

\section{Discussion}
\label{sec.Discussion}
In this section we provide an overview of our main results, highlighting the rigid definition of UGR tests established in this article, which can serve as a blueprint for future tests with atom-interferometric setups. 
Moreover, we roughly estimate the sensitivity to UGR violations of atom interferometer schemes, exemplified by the levitated geometry introduced in Fig.~\ref{fig:DifferentAIGeom}\hyperref[fig:DifferentAIGeom]{(b)} and of particular interest to state-of-the-art experimental implementations.
We compare it to other UGR-sensitive proposals, showing that this class of schemes may push the field towards more ambitious limits.

\subsection{Central results}
Idealized atomic clocks can be generated by steep quadratic potentials that force both internal states of an atom onto one common trajectory and colocalize them. 
The differential phase of two independent atomic clocks measures their proper-time difference, so that UGR violations may be detected.
In contrast, in Bragg-type atom interferometers, even when used for quantum clock interference, the interaction potential is usually linear and state independent.
As a consequence, in the Newtonian limit the phase only contains kinematic contributions, but the potential can be designed in such a way that these schemes mimic a UGR test.

Whereas most studies have so far focused on UGR tests solely generated by linear potentials introducing a momentum transfer~\cite{Loriani2019,Ufrecht2020}, we showed that quadratic or other nonlinear potentials can be used for atom-interferometric tests of UGR.
Therefore, guided interferometers operated with an internal superposition in principle give the same result as two independent atomic clocks.
Hence, they not only measure proper-time differences but also facilitate atom-interferometric tests of UGR.

However, a proper-time difference between the two branches of an interferometer is not necessary for tests of UGR, as highlighted in this work by the final two examples for atom interferometers with a variable mass.
It is evident that only proper-time differences between certain time segments of the interferometer are sufficient.
Consequently, the quest for interferometer geometries that display an intrinsic proper-time difference is not helpful~\cite{Wolf2010,Giese2019,Loriani2019,DiPumpo2020}.
A convenient choice of internal transitions seems more promising~\cite{Roura2020}.

Even internal superpositions as used for classical UGR tests with clocks are not essential~\cite{Ufrecht2020}.
Instead our results demonstrate that, for light-pulse schemes, internal transitions already enable the interferometric phase to include a UGR sensitivity, even without involving an internal superposition at any point.
As a consequence, a UGR sensitivity may also arise for Mach-Zehnder-type Raman geometries, including a Raman-based specular mirror interferometer~\cite{Giese2019}.
Taking the mass defect into account, one finds configurations that do not close in phase space. 
Such open geometries would introduce additional effects on the detected signal.
To this end, our treatment can be generalized by including open geometries and branch-dependent masses.

Finally, we derived a generalized treatment for both atomic clocks and atom interferometers, including physics beyond the Standard Model by considering non-Einsteinian components of gravity, thus closing the gap between these two subfields.
To this end, we demonstrated that the crucial mechanism for UGR tests, common to both clocks and atom interferometers, is a term involving $g$ times $\Delta m$ in the (differential) phase.
Furthermore, this term must also occur without the dilaton field and can be achieved, for example, through harmonic potentials.
Another possible mechanism includes a momentum change through an interaction that depends on the incoming internal state. 
Such interactions can be realized by state-dependent linear potentials or by a variable mass.
In addition, by introducing an additional acceleration that mimics the gravitational acceleration, Bragg-type atom interferometers can simulate the behavior of atomic clocks.
While the UGR sensitivity of such schemes was not discussed in previous works~\cite{Loriani2019,Roura2020,Ufrecht2020}, they highlight the impact of atom-interferometric UGR tests on fundamental tests of physics.

\subsection{Sensitivity estimation}
In order to obtain an estimate for the sensitivity of UGR tests, we recall that the discussed measurements of the gravitational redshift have the form $\Phi = -\Omega t_\text{red} (1+\alpha)\delta z_0 g /c^2 $, where the time $t_\text{red}$ is determined by the specific experiment and geometry.
In the following we assume shot-noise-limited measurements with $n_\text{at}$ atoms per shot and $T_\text{av}/t_\text{cyc}$ repetitions, where $T_\text{av}$ denotes the averaging time and $t_\text{cyc}$ the cycle time.
Hence, we find from Gaussian error propagation the uncertainty
\begin{equation}
    \label{eq:Deltalpha}
    \Delta \alpha = \left[\sqrt{n_\text{at}} \Omega \sqrt{\frac{t_\text{red}^2T_\text{av}}{t_\text{cyc}}} \frac{g \delta z_0}{c^2}\right]^{-1}
\end{equation}
for the violation parameter $\alpha$.

If we further assume that $t_\text{cyc} \sim t_\text{red}$, for fixed $T_\text{av}$, we obtain the uncertainty $\Delta \alpha \sim 1/(\delta z_0 \sqrt{t_\text{red}})$.
Hence, UGR tests benefit from an increasing spatial separation and, to a lesser degree, from increasing $t_\text{red}$.
This feature explains one of the big advantages of atomic clocks, where supreme distances $\delta z_0$ of several hundred meters or more~\cite{Delva2015,Herrmann2018,Delva2019,Savalle2019,Takamoto2020} are possible.
Indeed, analogous estimates for atomic clocks lead to uncertainties of $\Delta\alpha\sim 2.5\cdot 10^{-5}$~\cite{Delva2018,Delva2019} for space-based experiments and to $\Delta\alpha\sim 9\cdot 10^{-5}$~\cite{Takamoto2020} for earth-based setups.
However, the limitations of state-of-the-art atomic clocks are not solely given by this shot-noise-based estimate, but by other errors and noise sources.

Because the spatial separation in atom interferometers has to be generated within one experimental setup instead of performing two (independent) experiments at two locations, there is an intrinsic limitation on $\delta z_0$.
This limitation is even more relevant for UGR-sensitive atom interferometers in free fall based on internal transitions, as discussed in Sec.~\ref{sec.UGR_variable_mass}.
In this case, the dimensions of the apparatus limit both $\delta z_0$ and $t_\text{red}$ and are dictated by the duration of free fall in the experiment chamber.
Assuming a 10-m fountain, reasonable values for the schemes introduced in Ref.~\cite{Roura2020,Roura2020b} are $\delta z_0 \sim 1\,$cm and $t_\text{red}\sim 1\,$s.
For the scheme discussed in Ref.~\cite{Ufrecht2020}, the uncertainty of the violation parameter was estimated to reach $\Delta \alpha \sim 10^{-2}$ for strontium-88, about three orders of magnitude less sensitive than current atomic clock tests.

In contrast, compact devices with optical levitation suffer less severely from the limitations set by the spatial dimensions of the apparatus. Consequently, larger separations $\delta z_0$ and longer times $t_\text{red}$ are possible.
For example, semiguided schemes~\cite{Xu2019} can in principle achieve $t_\text{red}\sim 20\,$s, even though experiments demonstrating such a duration had a small spatial separation of $\delta z_0 \sim 10^{-4}\,$m.
However, they prove that realistic improvements of the uncertainty of $\alpha$ are within reach when resorting to mimicked UGR tests via levitating pulses akin to Fig.~\ref{fig:DifferentAIGeom}\hyperref[fig:DifferentAIGeom]{(b)}.
Indeed, we can estimate the order of magnitude of the signal of interest for such a scheme, assuming strontium-88 with $\Omega = 2\pi\times 429\,$THz and the effective two-photon Bragg wave vector $k_\text{eff}= 4 \pi /(813\,\text{nm})$~\cite{Katori2003}.
If we further assume large-momentum-transfer (LMT) beam splitters~\cite{Gebbe2021} with $400$ momentum transfers for the initial and final beam splitting, as well as fourth-order (magic) Bragg relaunching pulses, we find the spatial separation $\delta z_0 \sim 2\,$cm.
These parameters lead to a duration $t_\text{red}= N T \sim 9\,$s for a total of $N= 2000$ relaunch pulses.
Together with an averaging time of $T_\text{av}\sim 10^4\,$s and the atom number $n_\text{at}\sim 10^5$ we arrive at $\Delta \alpha \sim 10^{-3}$.
This result shows that levitated or guided schemes may improve the uncertainty about one order of magnitude compared to other UGR-sensitive proposals, reducing the gap to atomic clocks but still two orders of magnitude behind clock-based tests.
Of course, choosing different times for the initial separation (together with the associated duration of closing) and between the relaunch pulses in the setup of Fig.~\ref{fig:DifferentAIGeom}\hyperref[fig:DifferentAIGeom]{(b)} will increase $\delta z_0$ without sophisticated LMT techniques, improving $\Delta \alpha $ further.
However, the measured phase then includes additional terms that are also sensitive to UFF violations.
For a discussion of such additional terms, their experimental subtleties, as well as some error estimates, see Ref.~\cite{Ufrecht2020}.

Even though still limited to intermediate spatial separations, pulsed levitation or guided schemes in compact devices are a promising and important step to close the gap from atom interferometers to atomic clocks.
Besides, their intrinsically differential nature offers the possibility to mitigate some of the noise inherent to two independent setups.
Last but not least, there is a fundamental difference between such tests, since a single quantum object in both a spatial and internal superposition probes the basic principles of relativity, instead of two independent ones, as is the case for clocks. 
Therefore, levitated and guided schemes may represent an important contribution to quantum metrology and tests of fundamental physics.

\begin{acknowledgements}
We are grateful to T. Damour for his helpful support regarding the violation model. 
Moreover, we thank S. Loriani, E. Rasel, A. Roura, D. Schlippert, C. Schubert, and the QUANTUS team for fruitful and interesting discussions.
The project ``Metrology with interfering Unruh-DeWitt detectors'' (MIUnD) is funded by the Carl Zeiss Foundation (Carl-Zeiss-Stiftung).
The work of IQ\textsuperscript{ST} is financially supported by the Ministry of Science, Research and Art Baden-W\"urttemberg (Ministerium f\"ur Wissenschaft, Forschung und Kunst Baden-W\"urttemberg).
The QUANTUS and INTENTAS projects are supported by the German Aerospace Center (Deutsches Zentrum f\"ur Luft- und Raumfahrt, DLR) with funds provided by the Federal Ministry of Economic Affairs and Energy (Bundesministerium f\"ur Wirtschaft und Energie, BMWi) due to an enactment of the German Bundestag under Grant No. 50WM1956 (QUANTUS V) and No. 50WM2177-2178 (INTENTAS).
E.G. thanks the German Research Foundation (Deutsche Forschungsgemeinschaft, DFG) for a Mercator Fellowship within CRC 1227 (DQ-mat).
W.P.S. is grateful to Texas A\&M University for a Faculty Fellowship at the Hagler Institute for Advanced Study at Texas A\&M University and to Texas A\&M AgriLife for the support of this work.
W.G.U. acknowledges support by NSERC Canada (Natural  Science  and  Engineering  Research  Council), the Hagler Fellowship from HIAS (Hagler Institute for Advanced Study), the Department of Physics and the Institute for Quantum Science and Engineering (IQSE) at Texas A\&M University, as well as the Humboldt Foundation.
\end{acknowledgements}

\appendix

\section{Dilaton violation model}
\label{sec:DilMod}
The extension of the Standard Model by coupling to a scalar dilaton field $\varrho$ is motivated by string theory~\cite{Damour1994} as well as other approaches to quantum gravity. 
In these theories, the global coupling constant to gravity, the fermionic fields, and the gauge fields in the Lagrangian is a function of the dilaton field.
To separate this Lagrangian into a free part and an interacting part, it is convenient to perform a conformal transformation into the Einstein frame~\cite{Damour1994}.
In this frame the Lagrangian describing gravity appears as the conventional Einstein-Hilbert part together with additive modifications from the dilaton.
This way, an effective field theory can be derived. 
The coupling is given in this frame by the Lagrangian density $\mathcal{L}=\mathcal{L}_\text{free}+\mathcal{L}_\text{int}$ following Refs.~\cite{Damour1999,Damour2010}.
The first term of the free part of the theory
\begin{align}
\begin{split}
    \label{eq:DilIntLagFree}
    \mathcal{L}_\text{free}=&\frac{c^4}{16\pi G}\big[R-2(\nabla\varrho)^2\big]-\frac{1}{4\mu_0}F_{\mu\nu}F^{\mu\nu} \\
    &-\frac{1}{4}G^{\alpha}_{\mu\nu}G^{\mu\nu}_{\alpha}+\sum_{i=e,u,d}{\bar\psi_i\big[\ii\hbar c\slashed{D}-m_i c^2\big]\psi_i}
    \end{split}
\end{align}
includes the Ricci scalar $R$ and the kinetic term of the dilaton field $(\nabla\varrho)^2$, leading to the dilaton-modified Einstein field equations. 
Here, we introduced the (bare) Newtonian gravitational constant $G$. 
Furthermore, the Lagrangian contains the electromagnetic field with vacuum permeability $\mu_0$ and other gauge fields like the gluons with their respective field strength tensors $F_{\mu\nu}$ and $G_{\mu\nu}$. 
The last term describes the fermionic fields $\psi_i$ corresponding to electron, positron, up-quark and down-quark which make up ordinary matter. 
This simplified model does not include coupling to other quarks or the Higgs field and weak interaction. 
All derivatives and tensors in Eq.~\eqref{eq:DilIntLagFree} have to be understood as covariant constructs.

The interaction with the dilaton field is taken into account to first order in the dilaton. 
The Standard Model is therefore modified~\cite{Damour2010} through the interaction Lagrangian
\begin{align}
\begin{split}
\label{eq:DilIntLag}
        \mathcal{L}_\text{int}=&\varrho\left[\frac{d_e}{4\mu_0}F_{\mu\nu}F^{\mu\nu}-\frac{d_g\beta_3}{2 g_3}G^{\alpha}_{\mu\nu}G^{\mu\nu}_{\alpha}\right]  \\
    &-\varrho\sum_{i=e,u,d}{\bar\psi_i(d_{m_i}+\gamma_{m_i}d_g)m_i c^2\psi_i},
\end{split}
\end{align}
which parametrizes the dilaton coupling via the coefficients $d_s$, with $s = e, m_i, g$.
Like the coefficient $\gamma_{m_i}(g_3)$, the $\beta$-function $\beta_3(g_3)$ depends on the quantum-chromodynamics (QCD) coupling $g_3$, and both are defined as usual~\cite{Damour2010}.
In this effective model, we excluded the square root of the determinant of the general-relativistic metric tensor from the Lagrangian. 
Consequently, it appears in the spacetime volume element when integrating over this Lagrangian to obtain the action.

Combining Eqs.~\eqref{eq:DilIntLagFree} and~\eqref{eq:DilIntLag}, all coupling constants of the Standard Model become dilaton-dependent quantities.
In the case of fermions one defines $m_i(\varrho)=(1+d_{m_i}\varrho)m_i$ where $m_i$ is the (running) fermion mass calculated from renormalized QCD and the coupling $d_{m_i}$ is the renormalization-group-invariant modification caused by the dilaton.
As outlined in Ref.~\cite{Damour2010}, all dependence on the couplings $d_g, \beta_3$, as well as $g_3$ ---and thus to the QCD energy scale---has been absorbed into the (running) fermion masses $m_{i}$ as a consequence of the effective coarse-graining due to renormalization.
On the other hand, the coupling to the electromagnetic field is modified via $\mu_0(\varrho)\cong(1+d_e\varrho)\mu_0$. 
Here, $d_e$ again encodes the dilaton coupling of the electromagnetic field, which is equivalent to a dilaton-dependent fine-structure constant.

At this point one can treat bound systems like atoms within this field theory and arrives at dilaton-dependent energy levels $E_j(\varrho)$.
The explicit dependence of the energies $E_j$ on the dilaton field may be a highly non-trivial combination of all dilaton-coupling coefficients $d_\text{s}$ with $s=e,m_i,g$~\footnote{
If the focus lies for example on the optical transition spectrum in hydrogen-like atoms and by that on a specific energy scale, one can build a simpler effective theory by integrating out the energy scale associated with the nucleus.
This way, the problem reduces to an effective QED interaction between a positively charged atomic nucleus and a single shell electron.
In this new theory all dependence on the high energy scale is again absorbed into the coupling constants.}.
Hence, the mass $m_j(\varrho)=E_j(\varrho) / c^2$ associated with the energy of different internal states of an atom also depends on the dilaton field.
The non-relativistic limit is obtained by integrating out the relativistic degrees of freedom of the field theory in the spirit of Ref.~\cite{Caswell1986}.
In this approximation, the energy levels $E_j$ are subject to further corrections, like e.g. the Lamb-Shift or radiative corrections.
As a consequence of these dilaton-dependent coupling constants, we observe that a specific violation parameter measured either in UFF or in UGR tests depends on the exact composition of the nucleus as well as the electronic configuration of the atomic shell.
Hence, a comparison either of different rest masses or of different internal states leads to a different sensitivity to the fundamental dilaton couplings.
Moreover, it is possible to treat the interaction of the resulting theory with an external electromagnetic field via a non-relativistic effective field theory along the lines sketched in Ref.~\cite{Paz2015}.

Specifically, such a non-relativistic field theory~\cite{Pineda1998} can be matched to first-quantized models in the low-velocity and weak-field region where both are valid.
In this case the $m_j$ simply become the mass-energies of a first quantized description as introduced in Refs.~\cite{Anastopoulos18,Sonnleitner2018,Schwartz2019}. 
In case of a linear gravitational field this procedure leads to the Hamiltonian 
\begin{equation}
    \hat{H}^{(\sigma)}_j=m_j(\varrho)c^2+\frac{\hat{p}^2}{2m_j(\varrho)}+m_j(\varrho) g\hat z+\hat{V}_{j}^{(\sigma)}(\varrho),
\end{equation} 
which depends on the dilaton through two contributions:
The masses of the individual states and the effective potential $\hat{V}_{j}$ modelling the electromagnetic interaction.
After expanding the mass around its Standard-Model value via \smash{$m_j(\varrho)\cong (1+\bar\beta_j\varrho)m_j(0)$}, we find 
\begin{equation}
    \hat{H}^{(\sigma)}_j=(1+\bar\beta_j\varrho)m_j(0)c^2+\frac{\hat{p}^2}{2m_j(0)}+m_j(0) g\hat z+\hat{V}_{j}^{(\sigma)}(0).
\end{equation}
Here, the dependence on the dilaton is now only taken into account through the dominant term proportional to $c^2$.
Specifically, the interaction potential is independent of the dilaton field to this order.

Lastly, we stress that the dilaton field is not independent of the metric but can be understood as an additional source term in the Einstein field equations~\cite{Damour1994}.
This modification also manifests itself in the the weak-field limit and leads to the relation $\varrho=\bar\beta_\text{S} g z/c^2$.
The coupling coefficient $\bar\beta_\text{S}$ characterizes the source mass distribution responsible for the field.
Using this coefficient, we define an EEP-violating parameter $\beta_j=\bar\beta_j\bar\beta_\text{S}$.
In conclusion, this mechanisms allows us to parametrize EEP violations perturbatively in the low-velocity and weak-field limit by making the replacement $g\rightarrow(1+\beta_j)g$.

\section{Interference signal}
\label{sec:interference_signal}
The measured signal $I$ of a quantum interference experiment is the expectation value of a projection operator $\hat{\Pi}$.
This operator depends on the experiment and characterizes the measured observable, i.e., the output port of the interferometer.
Specifically, we find for an input state $\ket{\Psi_\text{in}}$ and the unitary time-evolution operator $\hat{U}$ the signal
\begin{equation}
    I = \bra{\Psi_\text{in}} \hat{U}^\dagger \hat{\Pi}\hat{U} \ket{\Psi_\text{in}} = \bra{\Psi_\text{in}} \hat{U}^\dagger_\Pi \hat{U}_\Pi^{\mathrlap{\phantom{\dagger_\Pi}}} \ket{\Psi_\text{in}}.
\end{equation}
In the last step, we have defined an effective time-evolution $\hat{U}_\Pi$ that postselects on the respective exit port~\cite{Kleinert2015} and consequently is not unitary.
The unitary operator $\hat{U}$ contains the complete interferometer sequence, including internal and c.m. degrees of freedom.

The typical experiments with atomic clocks and atom interferometers discussed in our article can be described in terms of the overlap of their c.m. wave-packet components. 
Hence, given an initial state of the form $\ket{\Psi_\text{in}} = \ket{\psi_\text{int}}\otimes \ket{\psi_\text{c.m.}}$ the interference signal can be written as
\begin{equation} \label{eq:interfernece}
    I = \frac{1}{4}  \bra{\psi_\text{c.m.}} (\hat{U}_1^\dagger+\hat{U}_2^\dagger)(\hat{U}_1+\hat{U}_2)\ket{\psi_\text{c.m.}} = \frac{1}{2} (1+C\cos \varphi),
\end{equation}
where the effective operators $\hat{U}_j$ solely act on the c.m. degree of freedom. 
With the second equal sign we introduced the contrast $C$ and phase $\varphi$ through the relation $ \bra{\psi_\text{c.m.}}\hat{U}^{\dagger}_1\hat{U}^{}_2\ket{\psi_\text{c.m.}}=C\exp({\ii\varphi})$.
Equation~\eqref{eq:interfernece} can be interpreted as the interference signal obtained from the superposition  $(\hat{U}_1+\hat{U}_2)\ket{\psi_\text{c.m.}}/2$ of two different components, one generated from the action of $\hat{U}_1$, the other from the action of $\hat{U}_2$.
In principle, one can associate with each of those operators an effective time evolution and by that an effective Hamiltonian.
Calculating the interference signal therefore reduces to taking the expectation value of the overlap operator $\hat{U}^{\dagger}_1\hat{U}^{}_2$ with respect to the initial state of the c.m. motion. 
This calculation can be performed, e.g., by suitable perturbative methods~\cite{Ufrecht2019,Ufrecht20202}.

When discussing atom interferometers, we conventionally measure the momentum of the atom at the end of the experiment. 
This measurement can be represented by the projection operator
\begin{equation}
    \hat \Pi =  \mathds{1}_\text{int} \otimes \int_{p_-}^{p_+} \mathrm{d} p \ket{p}\bra{p},
\end{equation}
where $p_-$ is the lower and $p_+$ the upper bound of the momentum interval that defines the exit port.
For the treatment of a Ramsey sequence used for atomic clocks we project on the superposition of internal states 
\begin{equation}
    \hat \Pi = \frac{\ket{a}+\ket{b}}{\sqrt{2}}\frac{\bra{a}+\bra{b}}{\sqrt{2}}   \otimes \mathds{1}_\text{ext}
\end{equation}
to model the final $\pi/2$ pulse and the read-out of the population in one internal state.

\section{Classical equations of motion}
\label{sec:ClassEqMot}
With the help of the classical, unperturbed counterpart of the Hamiltonian from Eq.~\eqref{eq:HamMassCorr}, we find the unperturbed equation of motion
\begin{equation}
    \label{eq:ClassEqOfMot}
    \ddot{z}^{(\sigma)}=-g+\frac{F^{(\sigma)}}{m}-\Gamma^2(z-\zeta^{(\sigma)})
\end{equation}
for atomic clocks and atom interferometers.
This equation of motion and the relations $\ddot{\bar z}=(\ddot{z}^{(u)}+\ddot{z}^{(l)})/2$ and \smash{$\big[\big(\dot{z}^{(u)}\big)^2-\big(\dot{z}^{(l)}\big)^2\big]/2=\dot{\bar z}\delta\dot{z}$} have been utilized to find the (differential) phases in the main part of the article. 
In this Appendix we will solve this equation for atomic clocks and light-pulse atom interferometers.

\subsection{Atomic clocks}
For atomic clocks we choose $F^{(\sigma)}=0$ and assume that the atoms are initially at rest and colocalized in the unperturbed potential minimum, i.e. $\dot{z}^{(\sigma)}(0)=0$ and $z^{(\sigma)}(0)=\zeta^{(\sigma)}(0)-g/\Gamma^2$.
In the following, we omit the superscript $\sigma$ for clarity.

For this choice the solution to Eq.~\eqref{eq:ClassEqOfMot} reads
\begin{equation}
\label{eq:exactSolution} 
    z(t)=\zeta(t)-\frac{g}{\Gamma^2}-\int_0^t\!\mathrm{d}t^\prime\,\mathrm{cos}(\Gamma[t-t^\prime])\dot{\zeta}(t^\prime).
\end{equation}
For an approximate solution in orders of $\Gamma^{-1}$, we obtain by successive application of partial integration
\begin{align}
\begin{split}
    z(t)=&\zeta(t)-\frac{1}{\Gamma^2}\big[g+\ddot{\zeta}(t)-\ddot{\zeta}(0)\cos{(\Gamma t)}\big] \\
    &+\frac{\dddot{\zeta}(0)}{\Gamma^3}\sin{(\Gamma t)}+\mathcal{O}(\Gamma^{-4}).
\end{split}
\end{align}
Here, we additionally assumed that the trap is initially at rest, that is $\dot{\zeta}(0)=0$.
Specifying further $\ddot{\bar\zeta}=0$, i.\,e. that the acceleration is chosen anti-symmetrically throughout the whole sequence, we find $\bar z-\bar\zeta=-g/\Gamma^2+\mathcal{O}(\Gamma^{-3})$ and $\delta z=\delta\zeta+\mathcal{O}(\Gamma^{-2})$.
Dropping this assumption would lead to an additional accelerational redshift~\cite{Okolow2020}.
For the geometry in Fig.~\ref{fig:DifferentClockGeom}\hyperref[fig:DifferentClockGeom]{(b)} the velocity $\dot{\zeta}=0$ and acceleration $\ddot{\zeta}=0$ vanishes, and $\bar z-\bar\zeta=-g/\Gamma^2$ as well as $\delta z=\delta\zeta_0$ are exact to the order considered here.

On the contrary, the scheme in Fig.~\ref{fig:DifferentClockGeom}\hyperref[fig:DifferentClockGeom]{(c)} consists of three different time segments, corresponding to the piecewise trajectories of the center of the trap
\begin{align}
\label{eq:trap_traj}
    \zeta(t)=\begin{cases}
    \zeta(0)+v t&T>t\geq 0 \\
    \zeta(0)+v T&T+T'>t\geq T \\
    \zeta(0)-v(t-[2T+T'])&2T+T'>t\geq T+T'.
\end{cases}
\end{align}
The lower branch is generated by replacing $v$ through $-v$.
With the help of Eq.~\eqref{eq:exactSolution} the trajectories of the atoms are easily obtained as
 $z(t)=\zeta(t)-g /\Gamma^2 -v \mathcal{S}(t) / \Gamma$, where the function
\begin{align}
   \mathcal{S}(t)=\begin{cases}
\sin{\Gamma t} \\
\sin{\Gamma t}-\sin{\big(\Gamma [t-T]\big)}\\
\sin{\Gamma t}-\sin{\big(\Gamma [t-T]\big)}-\sin{\big(\Gamma [t-(T+T')]\big)}
\end{cases}
\end{align}
is defined for the time segments in analogy to Eq.~\eqref{eq:trap_traj}.
We emphasize that the velocities are anti-symmetric for both branches, so the relations $\bar z-\bar\zeta=-g/\Gamma^2$ and $\delta z=\delta\zeta+\mathcal{O}(\Gamma^{-1})$ hold.

For the freely falling clocks in Fig.~\ref{fig:DifferentClockGeom}\hyperref[fig:DifferentClockGeom]{(a)}, we assume the same initial conditions but the trap is only turned on before $t=0$ with frequency $\Gamma(0)=\Gamma_0$.
After that it is turned off and the trajectory
\begin{equation}
    z(t)=\zeta(0)-g /\Gamma_0^2- g t^2/2
\end{equation}
describes a particle in free fall.
Hence, we obtain the results
$\dot{\bar z}(T)-\dot{\bar z}(0)=-gT$, as well as $\delta z=\delta\zeta_0=\zeta^{(u)}(0)-\zeta^{(l)}(0)$, which are both exact.

\subsection{Light-pulse atom interferometers}
We model light-pulse atom interferometers with a linear interaction potential, i.\,e., $\Gamma=0$.
The explicit force \smash{$F^{(\sigma)}=\hbar\sum_{\ell}k^{(\sigma)}_\ell\delta\left(t-t_\ell\right)$} of the instantaneous light pulses leads to a piecewise solution of the trajectory.
Therefore, we find for the trajectory
\begin{equation}
    z(t)=z(t_\ell)+\left[\dot{z}(t_\ell)+\frac{\hbar k_\ell}{m}\right]\left(t-t_\ell\right)-\frac{1}{2}g\left(t^2-t^2_\ell\right) 
\end{equation}
valid in the segment $t_{\ell+1}>t\geq t_\ell$, where we omitted the superscripts that label the branch.
Inserting a specific pulse sequence leads to the geometries discussed in the main part.
However, for guided atom interferometers we employ the same treatment used for atomic clocks, where no light pulses but a guiding potential is used.

Other linear interaction potentials for atom interferometers are also covered by Eq.~\eqref{eq:ClassEqOfMot}, but there the specific trajectory has to be calculated different from light pulses. 
In particular, if the unperturbed equation of motion involves a state-dependent force~\cite{Amit2019}, our treatment has to be modified.

\section{Wave-packet effects}
\label{sec:WPEffects}
Through a perturbative method~\cite{Ufrecht2019,Ufrecht20202} we derive for closed unperturbed geometries the phase 
\begin{equation}
\label{PerturbativeWPEffects}
    \varphi_\text{WP}=-\frac{1}{2\hbar}\oint\!\text{d}t\,\Big\lbrace\frac{\partial^2\mathcal{H}}{\partial z^2}\left<\hat{z}^2_0\right>+\frac{\partial^2\mathcal{H}}{\partial p^2}\left<\hat{p}^2_0\right> \Big\rbrace
\end{equation}
arising from wave-packet effects. 
In this equation no mixed derivatives appear since the perturbation Hamiltonian $\mathcal{H}$ in Eq.~\eqref{eq:PerturPot} includes no cross-term between momentum and position observables. 
Moreover, the centered time-dependent operators $\hat{z}_0(t)$ and $\hat{p}_0(t)$ have a vanishing expectation value \smash{$\left<\hat{z}_0\right>=\left<\hat{p}_0\right>=0$} and are defined below.
The integration path of the contour integral is defined along different masses for clocks or different branches for atom interferometers and corresponds to the definition of $\mathcal{H}_\text{diff}$ in the main part.

\subsection{Atomic clocks}
For atomic clocks we find from Eq.~\eqref{eq:PerturPot} the derivatives \smash{$\partial^2\mathcal{H}/\partial p^2=-\lambda_j\Delta m/(2m^2)$} and \smash{$\partial^2\mathcal{H}/\partial z^2=\lambda_j m\Delta\Gamma^2/2$}, if the trap is turned on for the whole sequence.
In addition, the centered observables can be calculated as \smash{$\hat{z}_0(t)=\big(\hat{z}-\left<\hat{z}\right>\big)\cos{(\Gamma t)}+\hat{p}\sin{(\Gamma t)}/(m\Gamma)$} and \smash{$\hat{p}_0(t)=\hat{p}\cos{(\Gamma t)}-m\Gamma\big(\hat{z}-\left<\hat{z}\right>\big)\sin{(\Gamma t)}$}, where we recall that we assumed a vanishing initial momentum $\left<\hat{p}\right>=0$ for atomic clocks.
With the help of these expressions we find for the phase originating from the wave-packet effects
\begin{equation}
    \varphi_\text{WP}=\frac{m\Gamma\Delta z^2(0)}{\hbar}\varphi_\text{zz}+\frac{\Delta p^2(0)}{\hbar m\Gamma}\varphi_\text{pp},
\end{equation}
where $\Delta z^2(0)$ as well as $\Delta p^2(0)$ are the branch-independent variances of the initial wave packet. 
For these quantities we require the conditions $m\Gamma\Delta z^2(0)/\hbar=\mathcal{O}(1)$ and $\Delta p^2(0)/(\hbar m\Gamma)=\mathcal{O}(1)$, i.\,e., that they scale with the variances of the ground state of the trap.
This way, we ensure that these terms cannot violate our perturbative treatment.
Additionally, we assumed that the expectation value of the cross term $\left<\hat{z}\hat{p}+\hat{p}\hat{z}\right>=0$  vanishes, which holds true for vanishing initial momenta and symmetric wave packets.
Finally, we obtain the branch-independent phase contributions
\begin{subequations}
\begin{equation}
    \varphi_\text{zz}=\frac{\Delta m}{m}\left[\frac{\Gamma T}{4}-\frac{\sin{(2\Gamma T)}}{8}\right]-\frac{\Delta\Gamma^2}{\Gamma^2}\left[\frac{\Gamma T}{4}+\frac{\sin{(2\Gamma T)}}{8}\right]
\end{equation}
and
\begin{equation}
    \varphi_\text{pp}=\frac{\Delta m}{m}\left[\frac{\Gamma T}{4}+\frac{\sin{(2\Gamma T)}}{8}\right]-\frac{\Delta\Gamma^2}{\Gamma^2}\left[\frac{\Gamma T}{4}-\frac{\sin{(2\Gamma T)}}{8}\right]
\end{equation}
\end{subequations}
where $T$ is the time the clock operates. We observe that for the perturbative treatment to be valid, to leading order the condition $\Delta\Gamma\ll\sqrt{\Gamma/T}$ has to be fulfilled. 

If the trap is turned off for times $t>0$, then we find the branch-independent phase
\begin{equation}
    \varphi_\text{WP}=\frac{\Delta m}{m}\frac{\Delta p^2(0)}{2\hbar m}T,
\end{equation}
arising only from different masses.
In all cases the wave-packet effects yield branch-independent phase contributions and consequently do not influence the differential phases.

\subsection{Closed atom interferometers with linear interaction potential}
For atom interferometers with a linear interaction potential including variable but branch-independent masses the derivatives of Eq.~\eqref{eq:PerturPot} are $\partial^2\mathcal{H}/\partial p^2=-\lambda(t)\Delta m/(2m^2)$ and $\partial^2\mathcal{H}/\partial z^2=0$.
The unperturbed equations of motion give rise to the centered observables $\hat{z}_0(t)=\hat{z}-\left<\hat{z}\right>+(\hat{p}-\left<\hat{p}\right>)t /m$ and $\hat{p}_0=\hat{p}-\left<\hat{p}\right>$.
Since these expressions yield a branch-independent integrand, the phase contribution $\varphi_\text{WP}$ from wave-packet effects vanishes for such interferometers, as already pointed out for example in Ref.~\cite{Loriani2019}.

\bibliography{RedshiftBib}

\end{document}